\documentclass[
    aps,
    prd,
    reprint,
    superscriptaddress,
    nolongbibliography,
    nobibnotes,
    nofootinbib,
]{revtex4-2}

\usepackage[dvipsnames]{xcolor}
\definecolor{linkcolor}{rgb}{0.0,0.3,0.5}

\usepackage[
    colorlinks=true,
    linkcolor=linkcolor,
    citecolor=linkcolor,
    filecolor=linkcolor,
    urlcolor=linkcolor,
    pdfusetitle,
]{hyperref}

\usepackage{amssymb}
\usepackage{amsmath}
\usepackage{physics}
\usepackage{orcidlink}

\usepackage{titlesec}
\titleformat{\section}[runin]{\itshape\bfseries}{}{0pt}{}[---]
\titlespacing{\section}{\parindent}{\parskip}{0pt}
\titleformat{\subsection}[runin]{\itshape}{}{0pt}{}[:]
\titlespacing{\subsection}{\parindent}{\parskip}{1ex}

\newcommand{\notts}{\affiliation{Nottingham Centre of Gravity \& School of Mathematical Sciences,\\University of Nottingham, University Park, Nottingham, NG7 2RD, United Kingdom}}
\newcommand{\milan}{\affiliation{Dipartimento di Fisica ``G. Occhialini'', Universit\'a degli Studi di Milano-Bicocca, Piazza della Scienza 3, 20126 Milano, Italy}}
\newcommand{\infn}{\affiliation{INFN, Sezione di Milano-Bicocca, Piazza della Scienza 3, 20126 Milano, Italy}}

\begin{document}

\title{Gravitational-wave astronomy requires population-informed parameter estimation}

\author{Matthew Mould$\,$\orcidlink{0000-0001-5460-2910}}
\email{matthew.mould@nottingham.ac.uk}
\notts

\author{Rodrigo Tenorio$\,$\orcidlink{0000-0002-3582-2587}}
\milan \infn

\author{Davide Gerosa$\,$\orcidlink{0000-0002-0933-3579}}
\milan \infn

\date{\today}

\begin{abstract}
Gravitational-wave events are interpreted in terms of Bayesian posteriors for their source properties inferred under unphysical reference priors. Though these parameter estimates are important intermediate data products for downstream analyses, we demonstrate that they are generically biased and therefore should not be used for astrophysical interpretation directly, as is common. Hierarchical parameter estimation is the solution, as joint analysis of the entire catalog of observations reduces statistical uncertainties and actually informs the correct prior, with population-informed event parameters now appropriate for astrophysical interpretation. As an example, we show how the most extreme measurements from a catalog can be derived and used to identify exceptional events from previous and ongoing observing runs, pointing out they are more informative about the population than any individual event. Using LIGO--Virgo--KAGRA data, we thus demonstrate that population inference is not optional to interpret gravitational-wave observations.
\end{abstract}

\maketitle

\section{Pitfalls of parameter estimation}

Gravitational-wave (GW) signals observed by the LIGO~\cite{LIGOScientific:2014pky}, Virgo~\cite{VIRGO:2014yos}, and KAGRA~\cite{KAGRA:2020tym}  (LVK) interferometers are used to measure the properties of their source compact-object mergers. Within a Bayesian framework, posterior distributions are inferred for each GW event independently~\cite{LIGOScientific:2018mvr, LIGOScientific:2020ibl, LIGOScientific:2021usb, KAGRA:2021vkt, LIGOScientific:2025slb, LIGOScientific:2026wfs}. The reference priors used are unphysical, in that they do not accurately represent the astrophysical distribution of sources (e.g., a uniform prior over detector-frame component masses), but are chosen to simplify downstream reuse of posterior samples~\cite{Ashton:2025xba}. However, this leads to generically biased parameter estimates that should be taken as intermediate data products only, as follows.

Consider the LVK parameter estimation (PE) pipeline, where a subset of sources from the astrophysical population are detected and analyzed individually. If the GW signal model, detector noise model, and inference algorithm are correct, the true source parameters should lie within the $P\,\%$ posterior interval $P\,\%$ of the time, but only if the prior matches the source distribution. Such ``perfect coverage'' is the key property of Bayesian inference that, explicitly or implicitly, is relied on to interpret data, as it guarantees the probability that source parameters fall within a given range of values. This coverage can be visualized using ``probability–probability’’ (P–P) plots---the cumulative distributions of the posterior levels at which the true source parameters fall---which should be diagonal (up to finite-sampling effects). This is necessary but not sufficient: meeting the condition does not guarantee correct posteriors, because other distributions also produce perfect coverage (e.g., the source distribution itself), but failure to does imply biases.

\begin{figure}
\centering
\includegraphics[width=1\columnwidth]{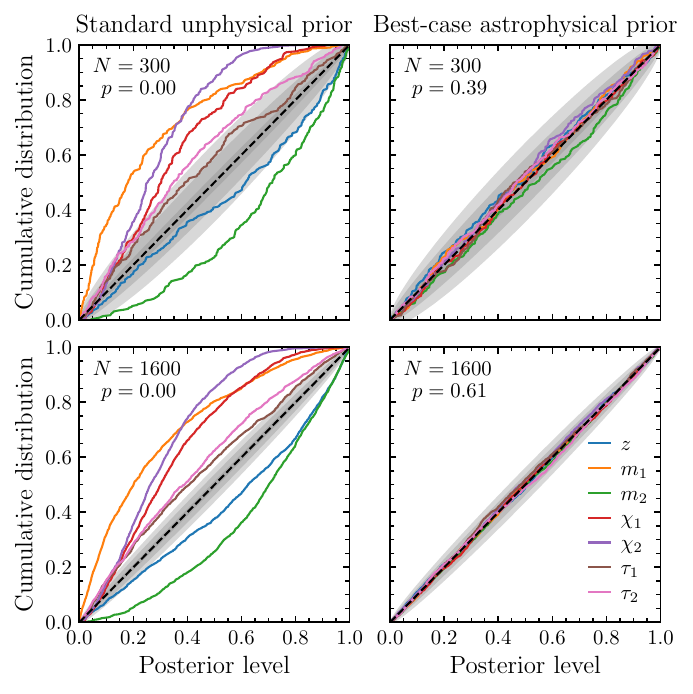}
\caption{P--P plot for 300 (top) and 1600 (bottom) detections simulated from an astrophysical population, analyzed with unphysical reference priors (left) or the most optimistic case of correct priors (right). Colored lines show cumulative distributions for the posterior levels at which the true redshifts $z$, primary and secondary source masses $m_1 \geq m_2$, spin magnitudes $\chi_{1,2}$, and spin--orbit tilt angles $\tau_{1,2}$ fall. Fisher-combined $p$-values for the expected uniform distribution (dashed black line) are given in each panel, with darker to lighter shaded gray regions showing $1\sigma$, $3\sigma$, and $5\sigma$ confidence intervals.}
\label{figure 1}
\end{figure}

As illustrated in Fig.~\ref{figure 1}, using the best-fit GWTC-3 population model~\cite{KAGRA:2021duu} from which we analyze detectable sources~\cite{Wolfe:2025yxu}, standard PE fails this test with significant biases. Comparing the expected and actual coverage with one-sample Kolmogorov--Smirnov (KS) tests yields $p$-values~$\approx0$. This is an unsurprising---yet perhaps underappreciated---conclusion, because the source distribution and Bayesian prior do not match. But crucially, it implies that standard PE results provide no guarantee on how accurately the source parameters are inferred, which can bias astrophysical interpretations~\cite{Vitale:2017cfs, Fishbach:2019ckx, Zevin:2020gxf, Mandel:2021ewy}.

\section{Hierarchical parameter estimation}

The solution is population inference with Bayesian hierarchical models~\cite{Loredo:2004nn, Mandel:2018mve, Vitale:2020aaz}. By analyzing the entire catalog of GW observations jointly, not only are statistical uncertainties reduced, but the observations themselves inform the correct astrophysical prior. At the same time, the individual source properties of all events are inferred again~\cite{KAGRA:2021duu, LIGOScientific:2025slb}, removing dependence on unphysical reference priors and alleviating posterior biases. Previous work has highlighted the importance of a population-level approach~\cite{Fishbach:2019ckx, Moore:2021xhn, vanHaasteren:2024yzz}, in particular on judging whether or not events are ``exceptional'' \cite{Fishbach:2019ckx, Essick:2021vlx, Mandel:2025qnh, Tenorio:2026dcc, Passenger:2024piv}.

Based only on single-event PE, the posterior for the source properties $\{\theta_n\}_{n=1}^N$ (masses, spins, etc.) of all $N$ independent events in the catalog is
\begin{align}
p_\mathrm{pe} ( \{\theta_n\} | \{d_n\} )
& =
\prod_{n=1}^N p_\mathrm{pe} ( \theta_n | d_n )
\, , \nonumber \\
p_\mathrm{pe} ( \theta_n | d_n )
& =
\frac
{ \mathcal{L}(d_n|\theta_n) p_\mathrm{pe}(\theta_n) }
{ \int \dd{\theta_n} \mathcal{L}(d_n|\theta_n) p_\mathrm{pe}(\theta_n) }
\, ,
\label{equation: joint pe}
\end{align}
using the shorthand notation $\{x_n\}$ for $\{x_n\}_{n=1}^N$. The usual GW likelihood is $\mathcal{L}(d|\theta)$~\cite{Thrane:2018qnx, Talbot:2025vth, LIGOScientific:2025hdt} and $p_\mathrm{pe}(\theta)$ is the unphysical reference prior that leads to biased posteriors.

Hierarchical PE instead gives the posterior
\begin{align}
p_\mathrm{pop} ( \lambda , \{\theta_n\} | \{d_n\} )
& =
p_\mathrm{pop} ( \lambda | \{d_n\} ) \prod_{n=1}^N
p_\mathrm{pop} ( \theta_n | d_n , \lambda )
\, , \nonumber \\
p_\mathrm{pop} ( \theta_n | d_n , \lambda )
& =
\frac
{ \mathcal{L} ( d_n | \theta_n ) p_\mathrm{pop} ( \theta_n | \lambda ) }
{ \int \dd{\theta_n} \mathcal{L} ( d_n | \theta_n ) p_\mathrm{pop} ( \theta_n | \lambda ) }
\, .
\label{equation: hierarchical pe}
\end{align}
The new prior $p_\mathrm{pop}(\theta|\lambda)$ is a population model with ``hyperparameters'' $\lambda$ that set the astrophysical distribution of source parameters $\theta$ and are inferred from the data, replacing $p_\mathrm{pe}(\theta)$. The posterior $p_\mathrm{pop}(\lambda|\{d_n\})$ for the hyperparameters accounts for GW selection effects~\cite{Mandel:2018mve, Vitale:2020aaz, LIGOScientific:2025pvj}. The so-called ``population-informed'' posterior $p_\mathrm{pop}(\{\theta_n\}|\{d_n\})$ represents our most accurate measurements for the source parameters of all events, replacing Eq.~\eqref{equation: joint pe}.

Though $p_\mathrm{pop}(\lambda|\{d_n\})$ is the typical target of population analyses~\cite{LIGOScientific:2018jsj, LIGOScientific:2020kqk, KAGRA:2021duu, LIGOScientific:2025pvj}, the joint posterior above can also be inferred, but is challenging due to the much higher dimensionality~\cite{Mancarella:2025uat}. Alternatively, given separate samples from $p_\mathrm{pop}(\lambda|\{d_n\})$ and from each $p_\mathrm{pe}(\theta_n|d_n)$, joint samples from $p_\mathrm{pop}( \lambda,\{\theta_n\}|\{d_n\})$ can be reconstructed as a postprocessing step~\cite{Fishbach:2019ckx, Galaudage:2019jdx, Miller:2020zox, Moore:2021xhn, T2100301, T1900895}. We describe this procedure and the hierarchical GW likelihood in full in Appendix~\ref{Hierarchical likelihood}.

\section{Impact on exceptional sources}

We reanalyze the population of black-hole (BH) mergers using LVK data through GWTC-4~\cite{LIGOScientific:2025slb}, constructing the hierarchical posterior in Eq.~\eqref{equation: hierarchical pe}. Our population prior is defined over source BH masses $m_1 \geq m_2$, spin magnitudes $\chi_{1,2}$, spin--orbit tilts $\tau_{1,2}$, and merger redshifts $z$. Full details of our analysis setup are given in Appendix~\ref{GWTC-4 analysis details}. Though hierarchical posteriors should always be used for astrophysical interpretation, we report comparisons with single-event PE to assess the impact of unphysical priors.

\begin{figure}
\centering
\includegraphics[width=1\columnwidth]{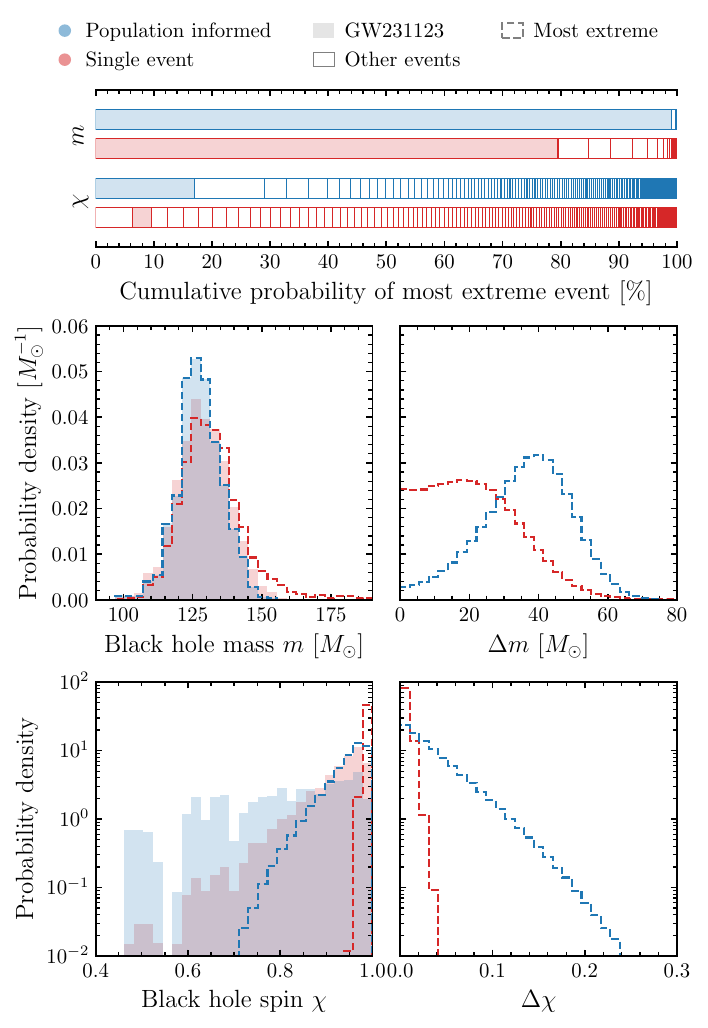}
\caption{Posterior distributions from single-event (red) and population-informed (blue) inference of GWTC-4. In the middle-left panel, dashed (shaded) histograms are for the mass $m$ of the heaviest BH in the catalog (primary BH in GW231123). The difference $\Delta m$ in mass between the two heaviest primary BHs in the catalog is shown in the middle-right panel. Dashed (shaded) histograms in the bottom-left panel are for the most rapid BH spin $\chi$ in the catalog (in GW231123), while the bottom-right panel shows the difference $\Delta\chi$ between the two most rapid BH spins in the catalog. The probabilities---ordered from highest to lowest---that each event has the most extreme BH mass or spin are shown in the top rows, with results for GW231123 shaded.}
\label{figure 2}
\end{figure}

In Fig.~\ref{figure 2}, we compare these posteriors for GW231123, as this event was highlighted as exceptional in GWTC-4 for having large BH masses and spins~\cite{LIGOScientific:2025rsn}. The inferred mass of its heavier (primary) BH is not significantly impacted by collective catalog information, but uncertainty is slightly reduced: $129_{-15}^{+16}M_\odot$ and $127_{-12}^{+14}M_\odot$ for single-event and population-informed PE, respectively (we report posterior medians and 90\% equal-tailed credible intervals throughout). Spin measurements are slightly more impacted: the population-informed spin of the more rapidly spinning BH~\cite{Biscoveanu:2020are} in GW231123, $0.84_{-0.24}^{+0.14}$, is less than that from single-event PE, $0.94_{-0.16}^{+0.04}$, but still high.

Given that this event was considered exceptional because of its source parameters, it can be compared to the most extreme parameter measurements across the catalog. These are found by ordering events according to a parameter of interest $\vartheta$, which could be one of the variables in $\theta$ (e.g., component BH mass) or derived from them (e.g., chirp mass), producing the so-called order statistics $\vartheta_{(1)} < \vartheta_{(2)} < ... < \vartheta_{(N)}$. Without loss of generality, the posterior for the most extreme source is found by selecting $\vartheta_{(N)} \equiv \max_n\{\vartheta_n\}$ for each sample from $p_\mathrm{pop}(\{\theta_n\}|\{d_n\})$. This is manifestly different from the usual selection based on, e.g., PE posterior median, as the above posterior mixes samples from different events. In particular, the probability that an event with index $i$ is the most extreme is found by recording $(N)\equiv\mathrm{argmax}_n\{\vartheta_n\}$ for each posterior sample and counting the number of times that $(N)=i$. While the usual approach derives a property of an individual event as informed by the catalog, our approach derives a property of the whole catalog.

The mass of the heaviest BH observed through GWTC-4 is $131_{-15}^{+23}M_\odot$ from single-event PE and $127_{-12}^{+14}M_\odot$ from population-informed PE, as shown in Fig.~\ref{figure 2}. In the former case, the heaviest BH is in the source of GW231123 with only 80\% posterior probability, but $\gtrsim99\%$ in the latter case. In fact, for single-event PE, the ``exceptionality'' of this event based on its total mass~\cite{LIGOScientific:2025rsn} is driven by the mass of its lighter (secondary) BH, which has $\gtrsim99\%$ posterior probability of being the heaviest secondary BH in the catalog in both cases.

As shown in Fig.~\ref{figure 2}, the highest BH spin measured through GWTC-4 is extreme, being $>0.98$ at 90\% credibility based on single-event PE. Population-informed PE also finds the highest spin to be near maximal, but with a less extreme constraint $>0.89$ at 90\% credibility. Unlike mass, there is no single event that has the highest BH spin in the catalog with dominant probability. Based on single-event PE, GW190517 (previously highlighted as having high spins~\cite{LIGOScientific:2020ibl}) has the highest probability with 6\%, compared to 3\% for GW231123; the order of these two events flips with population-informed PE, with probabilities 12\% and 17\%.

This procedure gives the most extreme source properties and the probability that each event attains this extremum, but not the degree of ``extremeness''. For example, one event may always be the most massive, but only by $0.1M_\odot$. This would not be considered exceptional, just a well constrained ``most extreme''. Complementary to other approaches~\cite{Fishbach:2019ckx, Essick:2021vlx}, order statistics provide a well-motivated definition of an exceptional event~\cite{Katz:2021cus}, according to $\Delta\vartheta := \vartheta_{(N)} - \vartheta_{(N-1)} \gg 0$, e.g., the most massive source is much more massive than the next.

In Fig.~\ref{figure 2}, we show the posterior for the difference $\Delta m$ between the two largest primary masses in the catalog. Single-event PE does not constrain $\Delta m$ away from zero, meaning the most massive BH in the catalog is not exceptional. Population-informed PE comes to the opposite conclusion, as the mass difference peaks at $\Delta m \approx 40M_\odot$ and is mostly constrained away from zero, with $\Delta m > 18 M_\odot$ at 90\% credibility. As GW231123 contains the most massive BH with high probability but its single-event and population-informed masses are similar, it instead is more exceptional after hierarchical analysis because the population-informed masses of other events are lower than their single-event counterparts.

 The difference between the two largest spins in Fig.~\ref{figure 2} is $\Delta\chi<0.01$ at 90\% credibility from single-event PE, but $\Delta\chi<0.08$ from population-informed PE. This implies there is no evidence for an exceptional BH spin, but the latter rules it out less strongly.

\begin{figure}
\centering
\includegraphics[width=1\columnwidth]{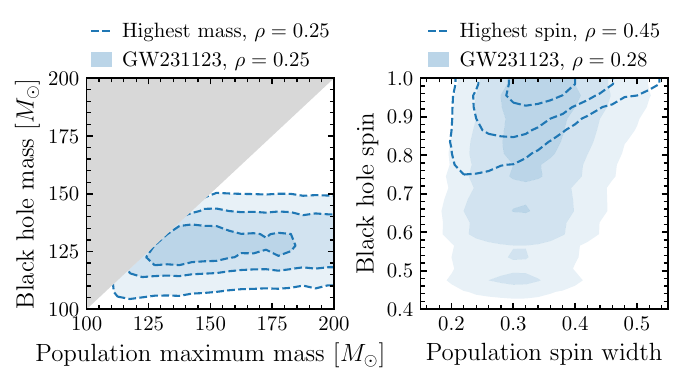}
\caption{Joint posteriors from GWTC-4 between select population parameters and the most extreme source parameters (dashed lines) or individual event parameters (shaded) they are most strongly correlated with, according to the Pearson correlation $\rho$. In each case, we show 50\%, 90\%, and 99\% credible regions. Left: the population maximum mass against the highest BH mass in the catalog and the primary mass of GW231123. Right: the width of the spin-magnitude population against the most rapid BH spin in the catalog and in GW231123.}
\label{figure 3}
\end{figure}

Note that Eq.~\eqref{equation: hierarchical pe} provides the posterior correlations of the population parameters $\lambda$ with source parameters $\{\theta_n\}$~\cite{Moore:2021xhn, Mancarella:2025uat} and thus also with the most extreme source parameter $\vartheta_{(N)}$---the property of the catalog that combines samples from different events. We show examples in Fig.~\ref{figure 3}. As the primary mass of GW231123 and the highest mass in the catalog are the same with high probability, they have equally strong correlations with the population maximum mass (Pearson correlation coefficients $\rho=0.25$). This is in contrast to the largest BH spin in the catalog, which is more correlated ($\rho=0.45$) with the extent of the spin-magnitude population than the highest spin in GW231123 ($\rho=0.28$), despite the latter being near maximal. In short, the population is necessarily most correlated with the most extreme parameter measurements, not individual events.

\section{Informing out-of-catalog events}

Interesting GW events in ongoing observing runs are highlighted before complete catalogs, e.g., GW241011 and GW241110~\cite{LIGOScientific:2025brd}. Though common~\cite{LIGOScientific:2020ufj, LIGOScientific:2024elc}, cherrypicked events should not be added to population fits~\cite{Essick:2021vlx}. Instead, the posterior
\begin{align}
p_\mathrm{pop}(\theta|d,\{d_n\})
\propto
\mathcal{L}(d|\theta)
p_\mathrm{pop}(\theta|\{d_n\})
\label{equation: informed event}
\end{align}
for the parameters $\theta$ of a new event with data $d$ can be informed independently by the existing catalog with the prior $p_\mathrm{pop}(\theta|\{d_n\}) = \int \dd{\lambda} p_\mathrm{pop}(\theta|\lambda) p_\mathrm{pop}(\lambda|\{d_n\})$. We assess the effect on the source properties of GW241011 and GW241110 using the population model above and a second model defined in terms of the effective spin~\cite{Racine:2008qv, LIGOScientific:2025hdt}
\begin{align}
\chi_\mathrm{eff} =
\frac { m_1 \chi_1 \cos\tau_1 + m_2 \chi_2 \cos\tau_2 } { m_1 + m_2 }
\end{align}
instead of $\chi_{1,2}$ and $\tau_{1,2}$, described in the Appendix~\ref{GWTC-4 analysis details}.

\begin{figure}
\centering
\includegraphics[width=1\columnwidth]{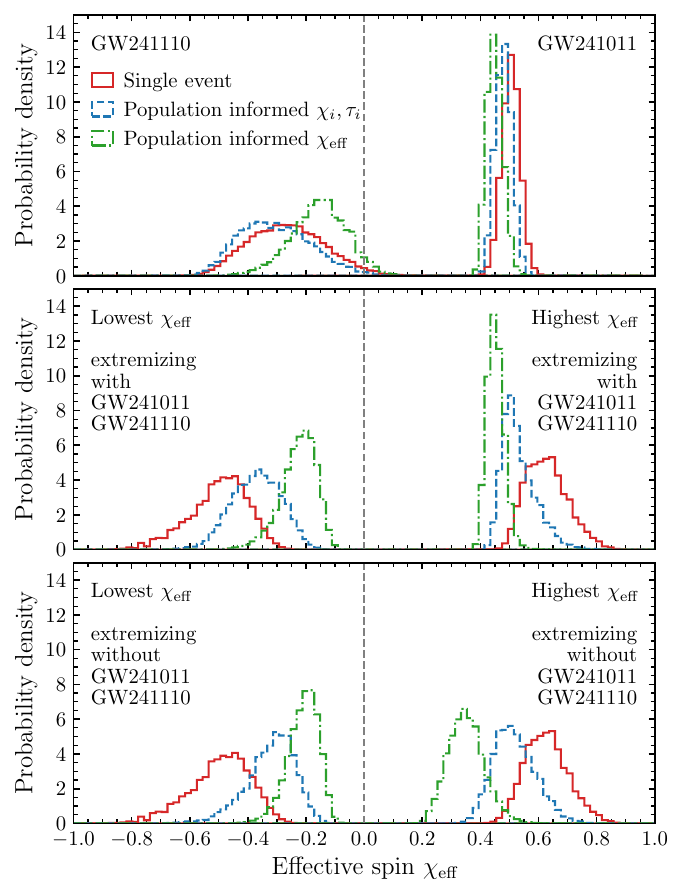}
\caption{Effective spin $\chi_\mathrm{eff}$ from single-event PE (red) and population-informed PE using population models defined in component spins $\chi_{1,2},\tau_{1,2}$ (blue) or $\chi_\mathrm{eff}$ (green). The top row shows posteriors for GW241011 (right) and GW241110 (left) informed by GWTC-4. The middle row shows the highest positive (right) and negative (left) $\chi_\mathrm{eff}$ in the catalog, including GW241011 and GW241110---when extremizing, not in the population fit itself; the bottom row shows the same without adding in those two events.}
\label{figure 4}
\end{figure}

We show the results in Fig.~\ref{figure 4}. Priors defined in component spins tend to produce posteriors with support for more extreme $\chi_\mathrm{eff}$ values, as these higher-dimensional spin degrees of freedom are less well measured than effective parameters and thus rule out less prior volume occupied by extreme sources. For GW241011, $\chi_\mathrm{eff}$ is well measured to be high from single-event PE ($\chi_\mathrm{eff}=0.51_{-0.05}^{+0.05}$), but is slightly lower from population-informed PE ($\chi_\mathrm{eff}=0.45_{-0.04}^{+0.05}$ with the $\chi_\mathrm{eff}$ population model). GW241110 is more poorly measured ($\chi_\mathrm{eff}=-0.27_{-0.20}^{+0.22}$ from single-event PE) and only influenced significantly by the $\chi_\mathrm{eff}$ population model ($\chi_\mathrm{eff}=-0.15_{-0.17}^{+0.15}$).

We also test if GW241011 and GW241110 inform the most extreme $\chi_\mathrm{eff}$ measurements---a hypothetical scenario with those events added to the catalog when selecting the most extreme parameter estimates, but not the population fit itself. Several events have support for high positive $\chi_\mathrm{eff}$, so there is only 5\% probability that GW241011 is the most extreme using single-event PE for all events; as shown in Fig.~\ref{figure 4}, the largest positive $\chi_\mathrm{eff}$ in the catalog is $0.62_{-0.10}^{+0.13}$, which is not changed by GW241011, despite its well-measured high $\chi_\mathrm{eff}$. In contrast, using the $\chi_\mathrm{eff}$ population model for population-informed PE of GWTC-4 events and correspondingly Eq.~\eqref{equation: informed event} for GW241011, the highest $\chi_\mathrm{eff}$ shifts from $0.35_{-0.09}^{+0.11}$ to $0.45_{-0.04}^{+0.05}$ and GW241011 has the highest $\chi_\mathrm{eff}$ with 93\% probability. The difference between the two highest $\chi_\mathrm{eff}$ measurements is not constrained away from zero ($<0.14$ at 90\% credibility), implying GW241011 does not have exceptionally high $\chi_\mathrm{eff}$. There are multiple events with support for negative $\chi_\mathrm{eff}$, so GW241110 is the most negative with only $<50\%$ probability across analyses in Fig.~\ref{figure 4}, and the most negative $\chi_\mathrm{eff}$ in the catalog is mostly unaffected by this event.

In all cases, population-informed PE is appreciably less extreme than single-event PE, highlighting that our knowledge of extreme sources in the catalog---those that test our astrophysical models most stringently---is unduly influenced by unphysical PE priors~\cite{Fishbach:2019ckx, Mandel:2025qnh, Tenorio:2026dcc}.

\section{Discussion}

Single-event PE and population inference are not disjoint analyses; in the context of Bayesian inference from catalog data, the latter is the correct approach to derive the former, not the other way round. This is equally true for ground-based~\cite{Fishbach:2019ckx, Moore:2021xhn}, spaced-based~\cite{Toubiana:2026yml, Criswell:2026xqk}, and pulsar-timing observations~\cite{vanHaasteren:2024yzz}. Recycling single-event PE samples for population inference is just a practical, operational recipe, but its accuracy is already limited at current catalog sizes~\cite{Farr:2019rap, Essick:2022ojx, Talbot:2023pex, Heinzel:2025ogf, Kobayashi:2026lac}. For astrophysical inference it may become necessary to characterize single-event likelihoods in more detail instead of posteriors~\cite{Loredo:2012jm} or bypass this step completely and sample the hierarchical posterior directly~\cite{Mancarella:2025uat, Criswell:2026xqk}.

Previous work has demonstrated the importance of hierarchical PE in judging apparent outliers~\cite{Fishbach:2019ckx, Mandel:2025qnh, Tenorio:2026dcc}. Our main conceptual point is further that any standard PE data product is inherently biased and interpretations thereof require more care than current GW catalogs emphasize. We illustrated this for a few among many interesting cases where unphysical PE priors are detrimental to astrophysical interpretation. We note that selecting events to study the population of BH mergers based on initial mass estimates is itself an incorrect use of single-event PE---an inconsistency present in all current analyses. Instead, all events should be included, with source classifications inferred hierarchically~\cite{Fishbach:2020ryj, Essick:2020ghc, LIGOScientific:2024elc}.

Though population models may be misspecified~\cite{Romero-Shaw:2022ctb, Miller:2026buq} just as unphysical PE priors are, they allow for much greater flexibility in specifying the source distribution and moreover are learnable from data. That standard PE priors do not make strong assumptions is a common misconception. Confirmation of new physics based on PE (e.g., orbital eccentricity~\cite{Romero-Shaw:2022xko, Gupte:2024jfe, Morras:2025xfu, Xu:2025ajj, Pompili:2026yxq} or effects beyond general relativity~\cite{Payne:2023kwj, Wu:2026qab}) will require careful judgment of these effects. Though single-event PE is less influenced by priors at high signal-to-noise ratios, our arguments are still relevant in this regime, as statements about the catalog as a whole depend on all events, not just the best measured few.

Going forward, only population-informed source parameter estimates derived from hierarchical catalog analyses should be used for concrete astrophysical interpretation of GW events.

\vspace{\baselineskip}
\section{Acknowledgments}

We thank
Noah Wolfe for providing data from Ref.~\cite{Wolfe:2025yxu} and for internal LIGO review,
Christopher Berry, Tristan Bruel, Katerina Chatziioannou, Reed Essick, Maya Fishbach, Jack Heinzel, and Simona Miller for helpful comments,
and the Sexten Center for Astrophysics where this work was kickstarted.
M.M. is supported by a Research Fellowship from the Royal Commission for the Exhibition of 1851.
R.T. and D.G. are supported by 
ERC Starting Grant No.~945155--GWmining, 
Cariplo Foundation Grant No.~2021-0555, 
Italian-French University (UIF/UFI) Grant No.~2025-C3-386,
MUR Grant ``Progetto Dipartimenti di Eccellenza 2023-2027'' (BiCoQ),
and the INFN TEONGRAV initiative.
D.G. is supported by MSCA Fellowship No.~101149270--ProtoBH and MUR Young Researchers Grant No.~SOE2024-0000125.
The authors are grateful for computational resources provided by
the LIGO Laboratory and supported by National Science Foundation Grants PHY-0757058 and PHY-0823459,
and CINECA with allocations through INFN and the University of Milano-Bicocca.
This research has made use of data or software obtained from the Gravitational Wave Open Science Center (gwosc.org), a service of the LIGO Scientific Collaboration, the Virgo Collaboration, and KAGRA,
and is based upon work supported by NSF's LIGO Laboratory, which is a major facility fully funded by the National Science Foundation.

\vspace{\baselineskip}
\section{Data availability}

The data that support the findings of this article are openly available~\cite{lvk-data, P2000223, ligo_scientific_collaboration_and_virgo_2022_6513631, ligo_scientific_collaboration_and_virgo_2023_8177023, ligo_scientific_collaboration_and_virgo_2025_17014085, ligo_scientific_collaboration_2025_16740128}.

\vspace{\baselineskip}
\appendix
\titleformat{\section}[runin]{\itshape\bfseries}{Appendix \thesection: }{0pt}{}[---]

\section{Hierarchical likelihood}
\label{Hierarchical likelihood}

The likelihood for the GW catalog can be derived as follows~\cite{Loredo:2004nn, Taylor:2018iat, Mandel:2018mve, Vitale:2020aaz, Essick:2023upv, Callister:2024cdx}. Assuming $N$ independent observations, the joint distribution of observed data $\{d_n\}_{n=1}^N$ and corresponding unobservable source parameters $\{\theta_n\}_{n=1}^N$ from an underlying population with hyperparameters $\lambda$ is
\begin{align}
p_\mathrm{pop}(\{d_n,\theta_n\}|\lambda) =
\prod_{n=1}^n p_\mathrm{pop}(d_n,\theta_n|\lambda,\mathrm{det})
\, .
\label{equation: product}
\end{align}
One must be careful to note that events included in the catalog are detectable by definition, made explicit by conditioning the contribution from each event on ``det.'' Writing the joint distribution in terms of independent events $p_\mathrm{pop}(\{d_n,\theta_n\}|\lambda) = \prod_{n=1}^N p_\mathrm{pop}(d_n,\theta_n|\lambda)$ is not correct, though not immediately apparent. The reason is a common abuse of notation, also used here, where ``$p_\mathrm{pop}$'' is used to denote both probability density functions in Eq.~\eqref{equation: product}. On the left-hand side, ``$p_\mathrm{pop}$'' already encapsulates that the catalog only contains detected events, but it does not on the right-hand side.

The observation of each event can be decomposed into three parts, written probabilistically as
\begin{align}
p_\mathrm{pop}(\mathrm{det},d,\theta|\lambda) =
P(\mathrm{det}|d) \mathcal{L}(d|\theta) p_\mathrm{pop}(\theta|\lambda)
\, .
\end{align}
As a generative model, this can be read from right to left: (i) a source with parameters $\theta$ is drawn from the underlying population distribution $p_\mathrm{pop}(\theta|\lambda)$, (ii) its signal is recorded in combination with detector noise as data $d$, represented by the likelihood $\mathcal{L}(d|\theta)$~\cite{Thrane:2018qnx, Talbot:2025vth, LIGOScientific:2025hdt}, and (iii) a search pipeline successfully detects the signal, where the detection probability is $P(\mathrm{det}|d_n)=1$ by definition for events in the catalog~\cite{Loredo:2004nn}. Using $p_\mathrm{pop}(\mathrm{det},d,\theta|\lambda) = p_\mathrm{pop}(d,\theta|\lambda,\mathrm{det}) P_\mathrm{pop}(\mathrm{det}|\lambda)$ and a prior $p_\mathrm{pop}(\lambda)$ gives
\begin{align}
p_\mathrm{pop} ( \lambda , \{\theta_n\} | \{d_n\} ) =
\frac { p_\mathrm{pop}(\lambda) } { p_\mathrm{pop}(\{d_n\}) }
\prod_{n=1}^N \frac
{ \mathcal{L}(d_n|\theta_n) p_\mathrm{pop}(\theta_n|\lambda) }
{ P_\mathrm{pop}(\mathrm{det}|\lambda) }
\label{equation: joint posterior}
\end{align}
for the joint posterior, where the fraction of sources from the underlying population that are detectable is
\begin{align}
P_\mathrm{pop}(\mathrm{det}|\lambda) = \int \dd{d} \dd{\theta}
P(\mathrm{det}|d) \mathcal{L}(d|\theta) p_\mathrm{pop}(\theta|\lambda)
\, .
\end{align}
The source parameters $\{\theta_n\}$ of individual events can be integrated out to obtain
\begin{align}
p_\mathrm{pop} ( \lambda | \{d_n\} ) =
\frac { p_\mathrm{pop}(\lambda) } { p_\mathrm{pop}(\{d_n\}) }
\prod_{n=1}^N \frac
{ p_\mathrm{pop}(d_n|\lambda) }
{ P_\mathrm{pop}(\mathrm{det}|\lambda) }
\, ,
\label{equation: population posterior}
\end{align}
where $p_\mathrm{pop}(d_n|\lambda) = \int \dd{\theta_n} \mathcal{L}(d_n|\theta_n) p_\mathrm{pop}(\theta_n|\lambda)$.

Comparing $p_\mathrm{pop}(\lambda,\{\theta_n\}|\{d_n\})$ and $p_\mathrm{pop}(\lambda|\{d_n\})$ immediately reveals that the former can be reconstructed from the latter, as is written in the main text. Then, note that Bayes' theorem implies a convenient rewriting:
\begin{align}
p_\mathrm{pop} ( \lambda , \{\theta_n\} | \{d_n\} ) =
p_\mathrm{pop}(\lambda|\{d_n\}) p_\mathrm{pe} ( \{\theta_n\} | \{d_n\} )
\nonumber \\
\times \prod_{n=1}^N \frac
{ p_\mathrm{pop}(\theta_n|\lambda) / p_\mathrm{pe}(\theta_n) }
{ \int\dd{\theta_n'} p_\mathrm{pe}(\theta_n'|d_n) p_\mathrm{pop}(\theta_n'|\lambda) / p_\mathrm{pe}(\theta_n') }
\, .
\label{equation: hierarchical resample}
\end{align}
Therefore, independent samples from $p_\mathrm{pop}(\lambda|\{d_n\})$ and $p_\mathrm{pe}(\{\theta_n\}|\{d_n\}$ can be reused as joint samples from $p_\mathrm{pop}(\lambda,\{\theta_n\}|\{d_n\})$ by weighing them according to Eq.~\eqref{equation: hierarchical resample}. A more efficient procedure is to first select one $\Lambda$ sample from $p_\mathrm{pop}(\Lambda|\{d_n\})$, then select samples $\theta_n$ from each $p_\mathrm{pe}(\theta_n|d_n)$ independently with probability $\propto p_\mathrm{pop}(\theta_n|\lambda) / p_\mathrm{pe}(\theta_n)$, then repeat many times~\cite{Miller:2020zox, Moore:2021xhn, T2100301, T1900895}.

\section{GWTC-4 analysis details}
\label{GWTC-4 analysis details}

We include the same 153 GW events confidently identified as binary BH mergers with false-alarm rates $< 1 \, \mathrm{yr}^{-1}$ as Ref.~\cite{LIGOScientific:2025pvj}. The GW signal models we use depend on their availability in the public LVK data~\cite{LIGOScientific:2019lzm, KAGRA:2023pio, LIGOScientific:2025snk}. For events in the latest GWTC-4 catalog~\cite{LIGOScientific:2025slb}, we use analyses with the \textsc{NRSur7dq4} waveform model~\cite{Varma:2019csw} if available, otherwise an equal weighting \textsc{IMRPhenomXPHM}~\cite{Pratten:2020ceb} and \textsc{SEOBNRv5PHM}~\cite{Ramos-Buades:2023ehm}. For events in GWTC-3~\cite{KAGRA:2021vkt}, we use the \textsc{IMRPhenomXPHM} and \textsc{SEOBNRv4PHM}~\cite{Ossokine:2020kjp} analyses, as these are the only ones available. For events in GWTC-1 through GWTC-2.1~\cite{LIGOScientific:2018mvr, LIGOScientific:2020ibl, LIGOScientific:2021usb}, we use the \textsc{NRSur7dq4} results from GWTC-2 if available, as this is the only catalog among those with \textsc{NRSur7dq4} analyses, otherwise \textsc{IMRPhenomXPHM} and \textsc{SEOBNRv4PHM} if both are available, or only \textsc{IMRPhenomXPHM} if not.

We use \textsc{gwax}~\cite{gwax} to compute the hierarchical GW likelihood. We estimate the $N+1$ integrals in Eq.~\eqref{equation: population posterior}---$p_\mathrm{pop}(d_n|\lambda)$ for $n=1,...,N$ and $P_\mathrm{pop}(\mathrm{det}|\lambda)$---with Monte Carlo integration, using each $p_\mathrm{pe}(\theta_n|d_n)$ and a distribution $q(\theta)$ from which signals are injected into detector noise and analyzed with search pipelines~\cite{Essick:2025zed}:
\begin{align}
& p_\mathrm{pop}(d_n|\lambda) \propto
\int \dd{\theta_n} p_\mathrm{pe}(\theta_n|d_n)
\frac { p_\mathrm{pop}(\theta|\lambda) } { p_\mathrm{pe}(\theta_n) }
\nonumber \\
& \approx
\frac { 1 } { S_n } \sum_{s=1}^{S_n}
\frac
{ p_\mathrm{pop}(\theta_{ns}|\lambda) }
{ p_\mathrm{pe}(\theta_{ns}) }
\, , \
\{ \theta_{ns} \}_{s=1}^{S_n} \sim p_\mathrm{pe}(\theta_n|d_n)
\, ;
\label{equation: mc likelihood}
\\
& P_\mathrm{pop}(\mathrm{det}|\lambda) =
\int \dd{d} \dd{\theta}
P(\mathrm{det}|d) \mathcal{L}(d|\theta) q(\theta)
\frac { p_\mathrm{pop}(\theta|\lambda) } { q(\theta) }
\nonumber \\
& \approx
\frac { 1 } { S } \sum_{s=1}^{S} P(\mathrm{det}|d_s)
\frac { p_\mathrm{pop}(\theta_s|\lambda) } { q(\theta_s) }
\, , \
\{d_s,\theta_s\}_{s=1}^S \sim \mathcal{L}(d|\theta) q(\theta)
\, .
\label{equation: mc pdet}
\end{align}
Note that these Monte Carlo estimates introduce additional uncertainty, resulting in an overall relative variance of the likelihood estimator~\cite{Tiwari:2017ndi, Farr:2019rap, Essick:2022ojx, Talbot:2023pex, Heinzel:2025ogf}. Following Ref.~\cite{LIGOScientific:2025pvj}, we estimate this relative variance using the same samples above and limit it to be less than unity, setting the likelihood to zero otherwise. We use \textsc{Bilby}~\cite{Ashton:2018jfp} to sample the population posterior with \textsc{Dynesty} nested sampling~\cite{Speagle:2019ivv}, using the settings \texttt{sample=`acceptance-walk'}, \texttt{naccept=10}, \texttt{nlive=500}.

The source parameters $\theta$ that we model with $p_\mathrm{pop}(\theta|\lambda)$ are the merger redshift $z$, source-frame BH masses $m_1 \geq m_2$, spin magnitudes $0 \leq \chi_{1,2} < 1$, and the spin--orbit tilt angles $0\leq\tau_{1,2}/\mathrm{rad}\leq\pi$. Our priors are similar but not identical to those in Ref.~\cite{LIGOScientific:2025pvj}, as follows:
\begin{itemize}
\item
The source number density over comoving volume $V$ and source-frame time evolves $\propto(1+z)^\gamma$, corresponding to a prior over redshift and detector-frame time $\propto(1+z)^{\gamma-1}\mathrm{d}V/\mathrm{d}z$~\cite{Fishbach:2018edt}. We assume a flat-$\Lambda$CDM cosmological model with Hubble parameter $H_0=67.9\,\mathrm{km}\,\mathrm{s}^{-1}\,\mathrm{Mpc}^{-1}$ and matter density $\Omega_\mathrm{m}=0.3065$~\cite{Planck:2015fie, LIGOScientific:2025hdt}.
\item
The prior on primary mass $m_1$ consists of a power-law distribution with slope $\alpha$ and two Gaussians with modes $\mu_{1,2}$ and scales $\sigma_{1,2}$, with mixing fractions $f_0$ and $f_{1,2}$, respectively. All three components are truncated to a maximum $m_\mathrm{max}$, a minimum $m_\mathrm{min}$, and are tapered at low mass by a sigmoid function $S(\frac{m_1-m_\mathrm{min}}{\delta_m})$ from $m_\mathrm{min}$ to $m_\mathrm{min} + \delta_m$, where $S(x) := x^2 (3-2x) H(x)$ and $H$ is the Heaviside step function. The product of this taper with power laws and Gaussians has closed-formed integral, such that the three tapered and truncated components of our $m_1$ distribution are normalized to unity in closed form (similar to Ref.~\cite{DeRenzis:2024dvx}).
\item
The population prior for $m_2$ is a power-law distribution with the same tapering and truncations as for $m_1$ (but by definition the maximum $m_2$ is $\min\{m_\mathrm{max},m_1\}$). This is exactly equivalent to a power-law distribution in the mass ratio $q=m_2/m_1$ with the same slope and low-mass taper.
\item
The population priors on both spin magnitudes $\chi_{1,2}$ are independent and identical Gaussians truncated on $[0,1)$ with mode $\mu_\chi$ and scale $\sigma_\chi$.
\item
For $\cos\tau_{1,2}$, we take independent and identical Gaussians truncated on $[-1,1]$ with mode $\mu_\tau$ and scale $\sigma_\tau$.
\item
For the model where we consider the effective spin $\chi_\mathrm{eff}$, we take a Gaussian truncated on $(-1,1)$ with mode $\mu_\mathrm{eff}$ and scale $\sigma_\mathrm{eff}$.
\end{itemize}
The remaining source parameters are not of astrophysical interest and our population priors for them match the original PE priors $p_\mathrm{pe}(\theta)$. The priors we take for the hyperparameters $\lambda$ of the above distributions are listed in Table~\ref{table: priors}.

\begin{table}
\caption{Priors on the hyperparameters of our population models. Intervals indicate ranges for uniform distributions and ``Dir.'' denotes a flat symmetric Dirichlet distribution.}
\centering
\setlength{\tabcolsep}{10pt}
\begin{tabular}{ccc}
\hline\hline
Symbol & Description & Range 
\\
\hline
$\gamma$ & $z$ power-law index & $[-10,10]$
\vspace{5pt} \\
$\beta$ & $m_2$ power-law index & $[-10,10]$
\\
$\alpha$ & $m_1$ power-law index & $[-10,10]$
\\
$\mu_1$ & Low $m_1$ peak $[M_\odot]$ & $[5,15]$
\\
$\mu_2$ & High $m_1$ peak $[M_\odot]$ & $[25,50]$
\\
$\sigma_1$ & Low $m_1$ peak width $[M_\odot]$ & $[0,10]$
\\
$\sigma_2$ & High $m_1$ peak width $[M_\odot]$ & $[0,10]$
\\
$f_{0,1,2}$ & Mixing fractions & Dir.
\\
$\delta_m$ & Low-mass smoothing $[M_\odot]$ & $[0,10]$
\\
$m_\mathrm{min}$ & Minimum mass $[M_\odot]$ & $[3,10]$
\\
$m_\mathrm{max}$ & Maximum mass $[M_\odot]$ & $[50,200]$
\vspace{5pt} \\
$\mu_\chi$ & Spin magnitude $\chi_{1,2}$ mode & $[0,1)$
\\
$\sigma_\chi$ & Spin magnitude $\chi_{1,2}$ width & $[0,1]$
\\
$\mu_\tau$ & Spin--orbit tilt $\cos\tau_{1,2}$ mode & $[-1,1]$
\\
$\sigma_\tau$ & Spin--orbit tilt $\cos\tau_{1,2}$ width & $[0,4]$
\vspace{5pt} \\
$\mu_\mathrm{eff}$ & Effective spin $\chi_\mathrm{eff}$ mode & $(-1,1)$
\\
$\sigma_\mathrm{eff}$ & Effective spin $\chi_\mathrm{eff}$ width & $[0,1]$
\\
\hline\hline
\end{tabular}
\label{table: priors}
\end{table}

\bibliography{draft}

\begin{thebibliography}{86}%
\makeatletter
\providecommand \@ifxundefined [1]{%
 \@ifx{#1\undefined}
}%
\providecommand \@ifnum [1]{%
 \ifnum #1\expandafter \@firstoftwo
 \else \expandafter \@secondoftwo
 \fi
}%
\providecommand \@ifx [1]{%
 \ifx #1\expandafter \@firstoftwo
 \else \expandafter \@secondoftwo
 \fi
}%
\providecommand \natexlab [1]{#1}%
\providecommand \enquote  [1]{``#1''}%
\providecommand \bibnamefont  [1]{#1}%
\providecommand \bibfnamefont [1]{#1}%
\providecommand \citenamefont [1]{#1}%
\providecommand \href@noop [0]{\@secondoftwo}%
\providecommand \href [0]{\begingroup \@sanitize@url \@href}%
\providecommand \@href[1]{\@@startlink{#1}\@@href}%
\providecommand \@@href[1]{\endgroup#1\@@endlink}%
\providecommand \@sanitize@url [0]{\catcode `\\12\catcode `\$12\catcode `\&12\catcode `\#12\catcode `\^12\catcode `\_12\catcode `\%12\relax}%
\providecommand \@@startlink[1]{}%
\providecommand \@@endlink[0]{}%
\providecommand \url  [0]{\begingroup\@sanitize@url \@url }%
\providecommand \@url [1]{\endgroup\@href {#1}{\urlprefix }}%
\providecommand \urlprefix  [0]{URL }%
\providecommand \Eprint [0]{\href }%
\providecommand \doibase [0]{https://doi.org/}%
\providecommand \selectlanguage [0]{\@gobble}%
\providecommand \bibinfo  [0]{\@secondoftwo}%
\providecommand \bibfield  [0]{\@secondoftwo}%
\providecommand \translation [1]{[#1]}%
\providecommand \BibitemOpen [0]{}%
\providecommand \bibitemStop [0]{}%
\providecommand \bibitemNoStop [0]{.\EOS\space}%
\providecommand \EOS [0]{\spacefactor3000\relax}%
\providecommand \BibitemShut  [1]{\csname bibitem#1\endcsname}%
\let\auto@bib@innerbib\@empty
\bibitem [{\citenamefont {Aasi}\ \emph {et~al.}(2015)\citenamefont {Aasi} \emph {et~al.}}]{LIGOScientific:2014pky}%
  \BibitemOpen
  \bibfield  {author} {\bibinfo {author} {\bibfnamefont {J.}~\bibnamefont {Aasi}} \emph {et~al.},\ }\href {https://doi.org/10.1088/0264-9381/32/7/074001} {\bibfield  {journal} {\bibinfo  {journal} {Class. Quantum Grav.}\ }\textbf {\bibinfo {volume} {32}},\ \bibinfo {pages} {074001} (\bibinfo {year} {2015})},\ \Eprint {https://arxiv.org/abs/1411.4547} {arXiv:1411.4547 [gr-qc]} \BibitemShut {NoStop}%
\bibitem [{\citenamefont {Acernese}\ \emph {et~al.}(2015)\citenamefont {Acernese} \emph {et~al.}}]{VIRGO:2014yos}%
  \BibitemOpen
  \bibfield  {author} {\bibinfo {author} {\bibfnamefont {F.}~\bibnamefont {Acernese}} \emph {et~al.},\ }\href {https://doi.org/10.1088/0264-9381/32/2/024001} {\bibfield  {journal} {\bibinfo  {journal} {Class. Quantum Grav.}\ }\textbf {\bibinfo {volume} {32}},\ \bibinfo {pages} {024001} (\bibinfo {year} {2015})},\ \Eprint {https://arxiv.org/abs/1408.3978} {arXiv:1408.3978 [gr-qc]} \BibitemShut {NoStop}%
\bibitem [{\citenamefont {Akutsu}\ \emph {et~al.}(2021)\citenamefont {Akutsu} \emph {et~al.}}]{KAGRA:2020tym}%
  \BibitemOpen
  \bibfield  {author} {\bibinfo {author} {\bibfnamefont {T.}~\bibnamefont {Akutsu}} \emph {et~al.},\ }\href {https://doi.org/10.1093/ptep/ptaa125} {\bibfield  {journal} {\bibinfo  {journal} {Prog. Theor. Exp. Phys.}\ }\textbf {\bibinfo {volume} {2021}},\ \bibinfo {pages} {05A101} (\bibinfo {year} {2021})},\ \Eprint {https://arxiv.org/abs/2005.05574} {arXiv:2005.05574 [physics.ins-det]} \BibitemShut {NoStop}%
\bibitem [{\citenamefont {Abbott}\ \emph {et~al.}(2019{\natexlab{a}})\citenamefont {Abbott} \emph {et~al.}}]{LIGOScientific:2018mvr}%
  \BibitemOpen
  \bibfield  {author} {\bibinfo {author} {\bibfnamefont {B.~P.}\ \bibnamefont {Abbott}} \emph {et~al.},\ }\href {https://doi.org/10.1103/PhysRevX.9.031040} {\bibfield  {journal} {\bibinfo  {journal} {Phys. Rev. X}\ }\textbf {\bibinfo {volume} {9}},\ \bibinfo {pages} {031040} (\bibinfo {year} {2019}{\natexlab{a}})},\ \Eprint {https://arxiv.org/abs/1811.12907} {arXiv:1811.12907 [astro-ph.HE]} \BibitemShut {NoStop}%
\bibitem [{\citenamefont {Abbott}\ \emph {et~al.}(2021{\natexlab{a}})\citenamefont {Abbott} \emph {et~al.}}]{LIGOScientific:2020ibl}%
  \BibitemOpen
  \bibfield  {author} {\bibinfo {author} {\bibfnamefont {R.}~\bibnamefont {Abbott}} \emph {et~al.},\ }\href {https://doi.org/10.1103/PhysRevX.11.021053} {\bibfield  {journal} {\bibinfo  {journal} {Phys. Rev. X}\ }\textbf {\bibinfo {volume} {11}},\ \bibinfo {pages} {021053} (\bibinfo {year} {2021}{\natexlab{a}})},\ \Eprint {https://arxiv.org/abs/2010.14527} {arXiv:2010.14527 [gr-qc]} \BibitemShut {NoStop}%
\bibitem [{\citenamefont {Abbott}\ \emph {et~al.}(2024)\citenamefont {Abbott} \emph {et~al.}}]{LIGOScientific:2021usb}%
  \BibitemOpen
  \bibfield  {author} {\bibinfo {author} {\bibfnamefont {R.}~\bibnamefont {Abbott}} \emph {et~al.},\ }\href {https://doi.org/10.1103/PhysRevD.109.022001} {\bibfield  {journal} {\bibinfo  {journal} {Phys. Rev. D}\ }\textbf {\bibinfo {volume} {109}},\ \bibinfo {pages} {022001} (\bibinfo {year} {2024})},\ \Eprint {https://arxiv.org/abs/2108.01045} {arXiv:2108.01045 [gr-qc]} \BibitemShut {NoStop}%
\bibitem [{\citenamefont {Abbott}\ \emph {et~al.}(2023{\natexlab{a}})\citenamefont {Abbott} \emph {et~al.}}]{KAGRA:2021vkt}%
  \BibitemOpen
  \bibfield  {author} {\bibinfo {author} {\bibfnamefont {R.}~\bibnamefont {Abbott}} \emph {et~al.},\ }\href {https://doi.org/10.1103/PhysRevX.13.041039} {\bibfield  {journal} {\bibinfo  {journal} {Phys. Rev. X}\ }\textbf {\bibinfo {volume} {13}},\ \bibinfo {pages} {041039} (\bibinfo {year} {2023}{\natexlab{a}})},\ \Eprint {https://arxiv.org/abs/2111.03606} {arXiv:2111.03606 [gr-qc]} \BibitemShut {NoStop}%
\bibitem [{\citenamefont {Abac}\ \emph {et~al.}(2025{\natexlab{a}})\citenamefont {Abac} \emph {et~al.}}]{LIGOScientific:2025slb}%
  \BibitemOpen
  \bibfield  {author} {\bibinfo {author} {\bibfnamefont {A.~G.}\ \bibnamefont {Abac}} \emph {et~al.},\ }\Eprint {https://arxiv.org/abs/2508.18082} {arXiv:2508.18082 [gr-qc]}  (\bibinfo {year} {2025}{\natexlab{a}})\BibitemShut {NoStop}%
\bibitem [{\citenamefont {Abac}\ \emph {et~al.}(2026)\citenamefont {Abac} \emph {et~al.}}]{LIGOScientific:2026wfs}%
  \BibitemOpen
  \bibfield  {author} {\bibinfo {author} {\bibfnamefont {A.~G.}\ \bibnamefont {Abac}} \emph {et~al.},\ }\Eprint {https://arxiv.org/abs/2605.27225} {arXiv:2605.27225 [gr-qc]}  (\bibinfo {year} {2026})\BibitemShut {NoStop}%
\bibitem [{\citenamefont {Ashton}(2026)}]{Ashton:2025xba}%
  \BibitemOpen
  \bibfield  {author} {\bibinfo {author} {\bibfnamefont {G.}~\bibnamefont {Ashton}},\ }\href {https://doi.org/10.1093/rasti/rzag012} {\bibfield  {journal} {\bibinfo  {journal} {RAS Tech. Instrum.}\ }\textbf {\bibinfo {volume} {5}},\ \bibinfo {pages} {rzag012} (\bibinfo {year} {2026})},\ \Eprint {https://arxiv.org/abs/2510.11197} {arXiv:2510.11197 [gr-qc]} \BibitemShut {NoStop}%
\bibitem [{\citenamefont {Abbott}\ \emph {et~al.}(2023{\natexlab{b}})\citenamefont {Abbott} \emph {et~al.}}]{KAGRA:2021duu}%
  \BibitemOpen
  \bibfield  {author} {\bibinfo {author} {\bibfnamefont {R.}~\bibnamefont {Abbott}} \emph {et~al.},\ }\href {https://doi.org/10.1103/PhysRevX.13.011048} {\bibfield  {journal} {\bibinfo  {journal} {Phys. Rev. X}\ }\textbf {\bibinfo {volume} {13}},\ \bibinfo {pages} {011048} (\bibinfo {year} {2023}{\natexlab{b}})},\ \Eprint {https://arxiv.org/abs/2111.03634} {arXiv:2111.03634 [astro-ph.HE]} \BibitemShut {NoStop}%
\bibitem [{\citenamefont {Wolfe}\ \emph {et~al.}(2025)\citenamefont {Wolfe}, \citenamefont {Mould}, \citenamefont {Heinzel},\ and\ \citenamefont {Vitale}}]{Wolfe:2025yxu}%
  \BibitemOpen
  \bibfield  {author} {\bibinfo {author} {\bibfnamefont {N.~E.}\ \bibnamefont {Wolfe}}, \bibinfo {author} {\bibfnamefont {M.}~\bibnamefont {Mould}}, \bibinfo {author} {\bibfnamefont {J.}~\bibnamefont {Heinzel}},\ and\ \bibinfo {author} {\bibfnamefont {S.}~\bibnamefont {Vitale}},\ }\Eprint {https://arxiv.org/abs/2510.06220} {arXiv:2510.06220 [gr-qc]}  (\bibinfo {year} {2025})\BibitemShut {NoStop}%
\bibitem [{\citenamefont {Vitale}\ \emph {et~al.}(2017)\citenamefont {Vitale}, \citenamefont {Gerosa}, \citenamefont {Haster}, \citenamefont {Chatziioannou},\ and\ \citenamefont {Zimmerman}}]{Vitale:2017cfs}%
  \BibitemOpen
  \bibfield  {author} {\bibinfo {author} {\bibfnamefont {S.}~\bibnamefont {Vitale}}, \bibinfo {author} {\bibfnamefont {D.}~\bibnamefont {Gerosa}}, \bibinfo {author} {\bibfnamefont {C.-J.}\ \bibnamefont {Haster}}, \bibinfo {author} {\bibfnamefont {K.}~\bibnamefont {Chatziioannou}},\ and\ \bibinfo {author} {\bibfnamefont {A.}~\bibnamefont {Zimmerman}},\ }\href {https://doi.org/10.1103/PhysRevLett.119.251103} {\bibfield  {journal} {\bibinfo  {journal} {Phys. Rev. Lett.}\ }\textbf {\bibinfo {volume} {119}},\ \bibinfo {pages} {251103} (\bibinfo {year} {2017})},\ \Eprint {https://arxiv.org/abs/1707.04637} {arXiv:1707.04637 [gr-qc]} \BibitemShut {NoStop}%
\bibitem [{\citenamefont {Fishbach}\ \emph {et~al.}(2020{\natexlab{a}})\citenamefont {Fishbach}, \citenamefont {Farr},\ and\ \citenamefont {Holz}}]{Fishbach:2019ckx}%
  \BibitemOpen
  \bibfield  {author} {\bibinfo {author} {\bibfnamefont {M.}~\bibnamefont {Fishbach}}, \bibinfo {author} {\bibfnamefont {W.~M.}\ \bibnamefont {Farr}},\ and\ \bibinfo {author} {\bibfnamefont {D.~E.}\ \bibnamefont {Holz}},\ }\href {https://doi.org/10.3847/2041-8213/ab77c9} {\bibfield  {journal} {\bibinfo  {journal} {Astrophys. J. Lett.}\ }\textbf {\bibinfo {volume} {891}},\ \bibinfo {pages} {L31} (\bibinfo {year} {2020}{\natexlab{a}})},\ \Eprint {https://arxiv.org/abs/1911.05882} {arXiv:1911.05882 [astro-ph.HE]} \BibitemShut {NoStop}%
\bibitem [{\citenamefont {Zevin}\ \emph {et~al.}(2020)\citenamefont {Zevin}, \citenamefont {Berry}, \citenamefont {Coughlin}, \citenamefont {Chatziioannou},\ and\ \citenamefont {Vitale}}]{Zevin:2020gxf}%
  \BibitemOpen
  \bibfield  {author} {\bibinfo {author} {\bibfnamefont {M.}~\bibnamefont {Zevin}}, \bibinfo {author} {\bibfnamefont {C.~P.~L.}\ \bibnamefont {Berry}}, \bibinfo {author} {\bibfnamefont {S.}~\bibnamefont {Coughlin}}, \bibinfo {author} {\bibfnamefont {K.}~\bibnamefont {Chatziioannou}},\ and\ \bibinfo {author} {\bibfnamefont {S.}~\bibnamefont {Vitale}},\ }\href {https://doi.org/10.3847/2041-8213/aba8ef} {\bibfield  {journal} {\bibinfo  {journal} {Astrophys. J. Lett.}\ }\textbf {\bibinfo {volume} {899}},\ \bibinfo {pages} {L17} (\bibinfo {year} {2020})},\ \Eprint {https://arxiv.org/abs/2006.11293} {arXiv:2006.11293 [astro-ph.HE]} \BibitemShut {NoStop}%
\bibitem [{\citenamefont {Mandel}\ and\ \citenamefont {Smith}(2021)}]{Mandel:2021ewy}%
  \BibitemOpen
  \bibfield  {author} {\bibinfo {author} {\bibfnamefont {I.}~\bibnamefont {Mandel}}\ and\ \bibinfo {author} {\bibfnamefont {R.~J.~E.}\ \bibnamefont {Smith}},\ }\href {https://doi.org/10.3847/2041-8213/ac35dd} {\bibfield  {journal} {\bibinfo  {journal} {Astrophys. J. Lett.}\ }\textbf {\bibinfo {volume} {922}},\ \bibinfo {pages} {L14} (\bibinfo {year} {2021})},\ \Eprint {https://arxiv.org/abs/2109.14759} {arXiv:2109.14759 [astro-ph.HE]} \BibitemShut {NoStop}%
\bibitem [{\citenamefont {Loredo}(2004)}]{Loredo:2004nn}%
  \BibitemOpen
  \bibfield  {author} {\bibinfo {author} {\bibfnamefont {T.~J.}\ \bibnamefont {Loredo}},\ }\href {https://doi.org/10.1063/1.1835214} {\bibfield  {journal} {\bibinfo  {journal} {AIP Conf. Proc.}\ }\textbf {\bibinfo {volume} {735}},\ \bibinfo {pages} {195} (\bibinfo {year} {2004})},\ \Eprint {https://arxiv.org/abs/astro-ph/0409387} {arXiv:astro-ph/0409387} \BibitemShut {NoStop}%
\bibitem [{\citenamefont {Mandel}\ \emph {et~al.}(2019)\citenamefont {Mandel}, \citenamefont {Farr},\ and\ \citenamefont {Gair}}]{Mandel:2018mve}%
  \BibitemOpen
  \bibfield  {author} {\bibinfo {author} {\bibfnamefont {I.}~\bibnamefont {Mandel}}, \bibinfo {author} {\bibfnamefont {W.~M.}\ \bibnamefont {Farr}},\ and\ \bibinfo {author} {\bibfnamefont {J.~R.}\ \bibnamefont {Gair}},\ }\href {https://doi.org/10.1093/mnras/stz896} {\bibfield  {journal} {\bibinfo  {journal} {Mon. Not. R. Astron. Soc.}\ }\textbf {\bibinfo {volume} {486}},\ \bibinfo {pages} {1086} (\bibinfo {year} {2019})},\ \Eprint {https://arxiv.org/abs/1809.02063} {arXiv:1809.02063 [physics.data-an]} \BibitemShut {NoStop}%
\bibitem [{\citenamefont {Vitale}\ \emph {et~al.}(2022)\citenamefont {Vitale}, \citenamefont {Gerosa}, \citenamefont {Farr},\ and\ \citenamefont {Taylor}}]{Vitale:2020aaz}%
  \BibitemOpen
  \bibfield  {author} {\bibinfo {author} {\bibfnamefont {S.}~\bibnamefont {Vitale}}, \bibinfo {author} {\bibfnamefont {D.}~\bibnamefont {Gerosa}}, \bibinfo {author} {\bibfnamefont {W.~M.}\ \bibnamefont {Farr}},\ and\ \bibinfo {author} {\bibfnamefont {S.~R.}\ \bibnamefont {Taylor}},\ }in\ \href {https://doi.org/10.1007/978-981-16-4306-4_45} {\emph {\bibinfo {booktitle} {Handbook of Gravitational Wave Astronomy}}}\ (\bibinfo  {publisher} {Springer},\ \bibinfo {address} {Singapore},\ \bibinfo {year} {2022})\ pp.\ \bibinfo {pages} {1709--1768},\ \Eprint {https://arxiv.org/abs/2007.05579} {arXiv:2007.05579 [astro-ph.IM]} \BibitemShut {NoStop}%
\bibitem [{\citenamefont {Moore}\ and\ \citenamefont {Gerosa}(2021)}]{Moore:2021xhn}%
  \BibitemOpen
  \bibfield  {author} {\bibinfo {author} {\bibfnamefont {C.~J.}\ \bibnamefont {Moore}}\ and\ \bibinfo {author} {\bibfnamefont {D.}~\bibnamefont {Gerosa}},\ }\href {https://doi.org/10.1103/PhysRevD.104.083008} {\bibfield  {journal} {\bibinfo  {journal} {Phys. Rev. D}\ }\textbf {\bibinfo {volume} {104}},\ \bibinfo {pages} {083008} (\bibinfo {year} {2021})},\ \Eprint {https://arxiv.org/abs/2108.02462} {arXiv:2108.02462 [gr-qc]} \BibitemShut {NoStop}%
\bibitem [{\citenamefont {van Haasteren}(2024)}]{vanHaasteren:2024yzz}%
  \BibitemOpen
  \bibfield  {author} {\bibinfo {author} {\bibfnamefont {R.}~\bibnamefont {van Haasteren}},\ }\href {https://doi.org/10.3847/1538-4365/ad530f} {\bibfield  {journal} {\bibinfo  {journal} {Astrophys. J. Supp. S.}\ }\textbf {\bibinfo {volume} {273}},\ \bibinfo {pages} {23} (\bibinfo {year} {2024})},\ \Eprint {https://arxiv.org/abs/2406.05081} {arXiv:2406.05081 [astro-ph.IM]} \BibitemShut {NoStop}%
\bibitem [{\citenamefont {Essick}\ \emph {et~al.}(2022)\citenamefont {Essick}, \citenamefont {Farah}, \citenamefont {Galaudage}, \citenamefont {Talbot}, \citenamefont {Fishbach}, \citenamefont {Thrane},\ and\ \citenamefont {Holz}}]{Essick:2021vlx}%
  \BibitemOpen
  \bibfield  {author} {\bibinfo {author} {\bibfnamefont {R.}~\bibnamefont {Essick}}, \bibinfo {author} {\bibfnamefont {A.}~\bibnamefont {Farah}}, \bibinfo {author} {\bibfnamefont {S.}~\bibnamefont {Galaudage}}, \bibinfo {author} {\bibfnamefont {C.}~\bibnamefont {Talbot}}, \bibinfo {author} {\bibfnamefont {M.}~\bibnamefont {Fishbach}}, \bibinfo {author} {\bibfnamefont {E.}~\bibnamefont {Thrane}},\ and\ \bibinfo {author} {\bibfnamefont {D.~E.}\ \bibnamefont {Holz}},\ }\href {https://doi.org/10.3847/1538-4357/ac3978} {\bibfield  {journal} {\bibinfo  {journal} {Astrophys. J.}\ }\textbf {\bibinfo {volume} {926}},\ \bibinfo {pages} {34} (\bibinfo {year} {2022})},\ \Eprint {https://arxiv.org/abs/2109.00418} {arXiv:2109.00418 [astro-ph.HE]} \BibitemShut {NoStop}%
\bibitem [{\citenamefont {Mandel}(2026)}]{Mandel:2025qnh}%
  \BibitemOpen
  \bibfield  {author} {\bibinfo {author} {\bibfnamefont {I.}~\bibnamefont {Mandel}},\ }\href {https://doi.org/10.3847/2041-8213/ae278d} {\bibfield  {journal} {\bibinfo  {journal} {Astrophys. J. Lett.}\ }\textbf {\bibinfo {volume} {996}},\ \bibinfo {pages} {L4} (\bibinfo {year} {2026})},\ \Eprint {https://arxiv.org/abs/2509.05885} {arXiv:2509.05885 [astro-ph.HE]} \BibitemShut {NoStop}%
\bibitem [{\citenamefont {Tenorio}\ and\ \citenamefont {Gerosa}(2026)}]{Tenorio:2026dcc}%
  \BibitemOpen
  \bibfield  {author} {\bibinfo {author} {\bibfnamefont {R.}~\bibnamefont {Tenorio}}\ and\ \bibinfo {author} {\bibfnamefont {D.}~\bibnamefont {Gerosa}},\ }\Eprint {https://arxiv.org/abs/2601.02467} {arXiv:2601.02467 [astro-ph.HE]}  (\bibinfo {year} {2026})\BibitemShut {NoStop}%
\bibitem [{\citenamefont {Passenger}\ \emph {et~al.}(2024)\citenamefont {Passenger}, \citenamefont {Thrane}, \citenamefont {Lasky}, \citenamefont {Payne}, \citenamefont {Stevenson},\ and\ \citenamefont {Farr}}]{Passenger:2024piv}%
  \BibitemOpen
  \bibfield  {author} {\bibinfo {author} {\bibfnamefont {L.}~\bibnamefont {Passenger}}, \bibinfo {author} {\bibfnamefont {E.}~\bibnamefont {Thrane}}, \bibinfo {author} {\bibfnamefont {P.~D.}\ \bibnamefont {Lasky}}, \bibinfo {author} {\bibfnamefont {E.}~\bibnamefont {Payne}}, \bibinfo {author} {\bibfnamefont {S.}~\bibnamefont {Stevenson}},\ and\ \bibinfo {author} {\bibfnamefont {B.}~\bibnamefont {Farr}},\ }\href {https://doi.org/10.1093/mnras/stae2521} {\bibfield  {journal} {\bibinfo  {journal} {Mon. Not. R. Astron. Soc.}\ }\textbf {\bibinfo {volume} {535}},\ \bibinfo {pages} {2837} (\bibinfo {year} {2024})},\ \Eprint {https://arxiv.org/abs/2405.09739} {arXiv:2405.09739 [astro-ph.HE]} \BibitemShut {NoStop}%
\bibitem [{\citenamefont {Thrane}\ and\ \citenamefont {Talbot}(2019)}]{Thrane:2018qnx}%
  \BibitemOpen
  \bibfield  {author} {\bibinfo {author} {\bibfnamefont {E.}~\bibnamefont {Thrane}}\ and\ \bibinfo {author} {\bibfnamefont {C.}~\bibnamefont {Talbot}},\ }\href {https://doi.org/10.1017/pasa.2019.2} {\bibfield  {journal} {\bibinfo  {journal} {Publ. Astron. Soc. Aust.}\ }\textbf {\bibinfo {volume} {36}},\ \bibinfo {pages} {e010} (\bibinfo {year} {2019})},\ \Eprint {https://arxiv.org/abs/1809.02293} {arXiv:1809.02293 [astro-ph.IM]} \BibitemShut {NoStop}%
\bibitem [{\citenamefont {Talbot}\ \emph {et~al.}(2025)\citenamefont {Talbot} \emph {et~al.}}]{Talbot:2025vth}%
  \BibitemOpen
  \bibfield  {author} {\bibinfo {author} {\bibfnamefont {C.}~\bibnamefont {Talbot}} \emph {et~al.},\ }\href {https://doi.org/10.1088/1361-6382/ae1ac7} {\bibfield  {journal} {\bibinfo  {journal} {Class. Quantum Grav.}\ }\textbf {\bibinfo {volume} {42}},\ \bibinfo {pages} {235023} (\bibinfo {year} {2025})},\ \Eprint {https://arxiv.org/abs/2508.11091} {arXiv:2508.11091 [gr-qc]} \BibitemShut {NoStop}%
\bibitem [{\citenamefont {Abac}\ \emph {et~al.}(2025{\natexlab{b}})\citenamefont {Abac} \emph {et~al.}}]{LIGOScientific:2025hdt}%
  \BibitemOpen
  \bibfield  {author} {\bibinfo {author} {\bibfnamefont {A.~G.}\ \bibnamefont {Abac}} \emph {et~al.},\ }\href {https://doi.org/10.3847/2041-8213/ae0c06} {\bibfield  {journal} {\bibinfo  {journal} {Astrophys. J. Lett.}\ }\textbf {\bibinfo {volume} {995}},\ \bibinfo {pages} {L18} (\bibinfo {year} {2025}{\natexlab{b}})},\ \Eprint {https://arxiv.org/abs/2508.18080} {arXiv:2508.18080 [gr-qc]} \BibitemShut {NoStop}%
\bibitem [{\citenamefont {Abac}\ \emph {et~al.}(2025{\natexlab{c}})\citenamefont {Abac} \emph {et~al.}}]{LIGOScientific:2025pvj}%
  \BibitemOpen
  \bibfield  {author} {\bibinfo {author} {\bibfnamefont {A.~G.}\ \bibnamefont {Abac}} \emph {et~al.},\ }\Eprint {https://arxiv.org/abs/2508.18083} {arXiv:2508.18083 [astro-ph.HE]}  (\bibinfo {year} {2025}{\natexlab{c}})\BibitemShut {NoStop}%
\bibitem [{\citenamefont {Abbott}\ \emph {et~al.}(2019{\natexlab{b}})\citenamefont {Abbott} \emph {et~al.}}]{LIGOScientific:2018jsj}%
  \BibitemOpen
  \bibfield  {author} {\bibinfo {author} {\bibfnamefont {B.~P.}\ \bibnamefont {Abbott}} \emph {et~al.},\ }\href {https://doi.org/10.3847/2041-8213/ab3800} {\bibfield  {journal} {\bibinfo  {journal} {Astrophys. J. Lett.}\ }\textbf {\bibinfo {volume} {882}},\ \bibinfo {pages} {L24} (\bibinfo {year} {2019}{\natexlab{b}})},\ \Eprint {https://arxiv.org/abs/1811.12940} {arXiv:1811.12940 [astro-ph.HE]} \BibitemShut {NoStop}%
\bibitem [{\citenamefont {Abbott}\ \emph {et~al.}(2021{\natexlab{b}})\citenamefont {Abbott} \emph {et~al.}}]{LIGOScientific:2020kqk}%
  \BibitemOpen
  \bibfield  {author} {\bibinfo {author} {\bibfnamefont {R.}~\bibnamefont {Abbott}} \emph {et~al.},\ }\href {https://doi.org/10.3847/2041-8213/abe949} {\bibfield  {journal} {\bibinfo  {journal} {Astrophys. J. Lett.}\ }\textbf {\bibinfo {volume} {913}},\ \bibinfo {pages} {L7} (\bibinfo {year} {2021}{\natexlab{b}})},\ \Eprint {https://arxiv.org/abs/2010.14533} {arXiv:2010.14533 [astro-ph.HE]} \BibitemShut {NoStop}%
\bibitem [{\citenamefont {Mancarella}\ and\ \citenamefont {Gerosa}(2025)}]{Mancarella:2025uat}%
  \BibitemOpen
  \bibfield  {author} {\bibinfo {author} {\bibfnamefont {M.}~\bibnamefont {Mancarella}}\ and\ \bibinfo {author} {\bibfnamefont {D.}~\bibnamefont {Gerosa}},\ }\href {https://doi.org/10.1103/PhysRevD.111.103012} {\bibfield  {journal} {\bibinfo  {journal} {Phys. Rev. D}\ }\textbf {\bibinfo {volume} {111}},\ \bibinfo {pages} {103012} (\bibinfo {year} {2025})},\ \Eprint {https://arxiv.org/abs/2502.12156} {arXiv:2502.12156 [gr-qc]} \BibitemShut {NoStop}%
\bibitem [{\citenamefont {Galaudage}\ \emph {et~al.}(2020)\citenamefont {Galaudage}, \citenamefont {Talbot},\ and\ \citenamefont {Thrane}}]{Galaudage:2019jdx}%
  \BibitemOpen
  \bibfield  {author} {\bibinfo {author} {\bibfnamefont {S.}~\bibnamefont {Galaudage}}, \bibinfo {author} {\bibfnamefont {C.}~\bibnamefont {Talbot}},\ and\ \bibinfo {author} {\bibfnamefont {E.}~\bibnamefont {Thrane}},\ }\href {https://doi.org/10.1103/PhysRevD.102.083026} {\bibfield  {journal} {\bibinfo  {journal} {Phys. Rev. D}\ }\textbf {\bibinfo {volume} {102}},\ \bibinfo {pages} {083026} (\bibinfo {year} {2020})},\ \Eprint {https://arxiv.org/abs/1912.09708} {arXiv:1912.09708 [astro-ph.HE]} \BibitemShut {NoStop}%
\bibitem [{\citenamefont {Miller}\ \emph {et~al.}(2020)\citenamefont {Miller}, \citenamefont {Callister},\ and\ \citenamefont {Farr}}]{Miller:2020zox}%
  \BibitemOpen
  \bibfield  {author} {\bibinfo {author} {\bibfnamefont {S.}~\bibnamefont {Miller}}, \bibinfo {author} {\bibfnamefont {T.~A.}\ \bibnamefont {Callister}},\ and\ \bibinfo {author} {\bibfnamefont {W.}~\bibnamefont {Farr}},\ }\href {https://doi.org/10.3847/1538-4357/ab80c0} {\bibfield  {journal} {\bibinfo  {journal} {Astrophys. J.}\ }\textbf {\bibinfo {volume} {895}},\ \bibinfo {pages} {128} (\bibinfo {year} {2020})},\ \Eprint {https://arxiv.org/abs/2001.06051} {arXiv:2001.06051 [astro-ph.HE]} \BibitemShut {NoStop}%
\bibitem [{\citenamefont {Callister}(2021)}]{T2100301}%
  \BibitemOpen
  \bibfield  {author} {\bibinfo {author} {\bibfnamefont {T.}~\bibnamefont {Callister}},\ }\href@noop {} {\bibinfo {title} {{LIGO DCC T2100301-v3}, \href{https://dcc.ligo.org/LIGO-T2100301/public}{dcc.ligo.org/LIGO-T2100301/public}}} (\bibinfo {year} {2021})\BibitemShut {NoStop}%
\bibitem [{\citenamefont {Essick}\ and\ \citenamefont {Fishbach}(2021)}]{T1900895}%
  \BibitemOpen
  \bibfield  {author} {\bibinfo {author} {\bibfnamefont {R.}~\bibnamefont {Essick}}\ and\ \bibinfo {author} {\bibfnamefont {M.}~\bibnamefont {Fishbach}},\ }\href@noop {} {\bibinfo {title} {{LIGO DCC T1900895-v2}, \href{https://dcc.ligo.org/LIGO-T1900895/public}{dcc.ligo.org/LIGO-T1900895/public}}} (\bibinfo {year} {2021})\BibitemShut {NoStop}%
\bibitem [{\citenamefont {Abac}\ \emph {et~al.}(2025{\natexlab{d}})\citenamefont {Abac} \emph {et~al.}}]{LIGOScientific:2025rsn}%
  \BibitemOpen
  \bibfield  {author} {\bibinfo {author} {\bibfnamefont {A.~G.}\ \bibnamefont {Abac}} \emph {et~al.},\ }\href {https://doi.org/10.3847/2041-8213/ae0c9c} {\bibfield  {journal} {\bibinfo  {journal} {Astrophys. J. Lett.}\ }\textbf {\bibinfo {volume} {993}},\ \bibinfo {pages} {L25} (\bibinfo {year} {2025}{\natexlab{d}})},\ \Eprint {https://arxiv.org/abs/2507.08219} {arXiv:2507.08219 [astro-ph.HE]} \BibitemShut {NoStop}%
\bibitem [{\citenamefont {Biscoveanu}\ \emph {et~al.}(2021)\citenamefont {Biscoveanu}, \citenamefont {Isi}, \citenamefont {Vitale},\ and\ \citenamefont {Varma}}]{Biscoveanu:2020are}%
  \BibitemOpen
  \bibfield  {author} {\bibinfo {author} {\bibfnamefont {S.}~\bibnamefont {Biscoveanu}}, \bibinfo {author} {\bibfnamefont {M.}~\bibnamefont {Isi}}, \bibinfo {author} {\bibfnamefont {S.}~\bibnamefont {Vitale}},\ and\ \bibinfo {author} {\bibfnamefont {V.}~\bibnamefont {Varma}},\ }\href {https://doi.org/10.1103/PhysRevLett.126.171103} {\bibfield  {journal} {\bibinfo  {journal} {Phys. Rev. Lett.}\ }\textbf {\bibinfo {volume} {126}},\ \bibinfo {pages} {171103} (\bibinfo {year} {2021})},\ \Eprint {https://arxiv.org/abs/2007.09156} {arXiv:2007.09156 [astro-ph.HE]} \BibitemShut {NoStop}%
\bibitem [{\citenamefont {Katz}(2021)}]{Katz:2021cus}%
  \BibitemOpen
  \bibfield  {author} {\bibinfo {author} {\bibfnamefont {J.~I.}\ \bibnamefont {Katz}},\ }\href {https://doi.org/10.1093/mnras/stab2551} {\bibfield  {journal} {\bibinfo  {journal} {Mon. Not. R. Astron. Soc.}\ }\textbf {\bibinfo {volume} {508}},\ \bibinfo {pages} {69} (\bibinfo {year} {2021})},\ \Eprint {https://arxiv.org/abs/2106.05212} {arXiv:2106.05212 [astro-ph.HE]} \BibitemShut {NoStop}%
\bibitem [{\citenamefont {Abac}\ \emph {et~al.}(2025{\natexlab{e}})\citenamefont {Abac} \emph {et~al.}}]{LIGOScientific:2025brd}%
  \BibitemOpen
  \bibfield  {author} {\bibinfo {author} {\bibfnamefont {A.~G.}\ \bibnamefont {Abac}} \emph {et~al.},\ }\href {https://doi.org/10.3847/2041-8213/ae0d54} {\bibfield  {journal} {\bibinfo  {journal} {Astrophys. J. Lett.}\ }\textbf {\bibinfo {volume} {993}},\ \bibinfo {pages} {L21} (\bibinfo {year} {2025}{\natexlab{e}})},\ \Eprint {https://arxiv.org/abs/2510.26931} {arXiv:2510.26931 [astro-ph.HE]} \BibitemShut {NoStop}%
\bibitem [{\citenamefont {Abbott}\ \emph {et~al.}(2020)\citenamefont {Abbott} \emph {et~al.}}]{LIGOScientific:2020ufj}%
  \BibitemOpen
  \bibfield  {author} {\bibinfo {author} {\bibfnamefont {R.}~\bibnamefont {Abbott}} \emph {et~al.},\ }\href {https://doi.org/10.3847/2041-8213/aba493} {\bibfield  {journal} {\bibinfo  {journal} {Astrophys. J. Lett.}\ }\textbf {\bibinfo {volume} {900}},\ \bibinfo {pages} {L13} (\bibinfo {year} {2020})},\ \Eprint {https://arxiv.org/abs/2009.01190} {arXiv:2009.01190 [astro-ph.HE]} \BibitemShut {NoStop}%
\bibitem [{\citenamefont {Abac}\ \emph {et~al.}(2024)\citenamefont {Abac} \emph {et~al.}}]{LIGOScientific:2024elc}%
  \BibitemOpen
  \bibfield  {author} {\bibinfo {author} {\bibfnamefont {A.~G.}\ \bibnamefont {Abac}} \emph {et~al.},\ }\href {https://doi.org/10.3847/2041-8213/ad5beb} {\bibfield  {journal} {\bibinfo  {journal} {Astrophys. J. Lett.}\ }\textbf {\bibinfo {volume} {970}},\ \bibinfo {pages} {L34} (\bibinfo {year} {2024})},\ \Eprint {https://arxiv.org/abs/2404.04248} {arXiv:2404.04248 [astro-ph.HE]} \BibitemShut {NoStop}%
\bibitem [{\citenamefont {Racine}(2008)}]{Racine:2008qv}%
  \BibitemOpen
  \bibfield  {author} {\bibinfo {author} {\bibfnamefont {E.}~\bibnamefont {Racine}},\ }\href {https://doi.org/10.1103/PhysRevD.78.044021} {\bibfield  {journal} {\bibinfo  {journal} {Phys. Rev. D}\ }\textbf {\bibinfo {volume} {78}},\ \bibinfo {pages} {044021} (\bibinfo {year} {2008})},\ \Eprint {https://arxiv.org/abs/0803.1820} {arXiv:0803.1820 [gr-qc]} \BibitemShut {NoStop}%
\bibitem [{\citenamefont {Toubiana}\ and\ \citenamefont {Gair}(2026)}]{Toubiana:2026yml}%
  \BibitemOpen
  \bibfield  {author} {\bibinfo {author} {\bibfnamefont {A.}~\bibnamefont {Toubiana}}\ and\ \bibinfo {author} {\bibfnamefont {J.}~\bibnamefont {Gair}},\ }\Eprint {https://arxiv.org/abs/2601.04168} {arXiv:2601.04168 [gr-qc]}  (\bibinfo {year} {2026})\BibitemShut {NoStop}%
\bibitem [{\citenamefont {Criswell}\ \emph {et~al.}(2026)\citenamefont {Criswell}, \citenamefont {Banagiri}, \citenamefont {Delfavero}, \citenamefont {Bustamante-Rosell}, \citenamefont {Taylor},\ and\ \citenamefont {Rosati}}]{Criswell:2026xqk}%
  \BibitemOpen
  \bibfield  {author} {\bibinfo {author} {\bibfnamefont {A.~W.}\ \bibnamefont {Criswell}}, \bibinfo {author} {\bibfnamefont {S.}~\bibnamefont {Banagiri}}, \bibinfo {author} {\bibfnamefont {V.}~\bibnamefont {Delfavero}}, \bibinfo {author} {\bibfnamefont {M.~J.}\ \bibnamefont {Bustamante-Rosell}}, \bibinfo {author} {\bibfnamefont {S.~R.}\ \bibnamefont {Taylor}},\ and\ \bibinfo {author} {\bibfnamefont {R.}~\bibnamefont {Rosati}},\ }\Eprint {https://arxiv.org/abs/2604.03390} {arXiv:2604.03390 [astro-ph.IM]}  (\bibinfo {year} {2026})\BibitemShut {NoStop}%
\bibitem [{\citenamefont {Farr}(2019)}]{Farr:2019rap}%
  \BibitemOpen
  \bibfield  {author} {\bibinfo {author} {\bibfnamefont {W.~M.}\ \bibnamefont {Farr}},\ }\href {https://doi.org/10.3847/2515-5172/ab1d5f} {\bibfield  {journal} {\bibinfo  {journal} {Res. Notes Am. Astron. Soc.}\ }\textbf {\bibinfo {volume} {3}},\ \bibinfo {pages} {66} (\bibinfo {year} {2019})},\ \Eprint {https://arxiv.org/abs/1904.10879} {arXiv:1904.10879 [astro-ph.IM]} \BibitemShut {NoStop}%
\bibitem [{\citenamefont {Essick}\ and\ \citenamefont {Farr}(2022)}]{Essick:2022ojx}%
  \BibitemOpen
  \bibfield  {author} {\bibinfo {author} {\bibfnamefont {R.}~\bibnamefont {Essick}}\ and\ \bibinfo {author} {\bibfnamefont {W.}~\bibnamefont {Farr}},\ }\Eprint {https://arxiv.org/abs/2204.00461} {arXiv:2204.00461 [astro-ph.IM]}  (\bibinfo {year} {2022})\BibitemShut {NoStop}%
\bibitem [{\citenamefont {Talbot}\ and\ \citenamefont {Golomb}(2023)}]{Talbot:2023pex}%
  \BibitemOpen
  \bibfield  {author} {\bibinfo {author} {\bibfnamefont {C.}~\bibnamefont {Talbot}}\ and\ \bibinfo {author} {\bibfnamefont {J.}~\bibnamefont {Golomb}},\ }\href {https://doi.org/10.1093/mnras/stad2968} {\bibfield  {journal} {\bibinfo  {journal} {Mon. Not. R. Astron. Soc.}\ }\textbf {\bibinfo {volume} {526}},\ \bibinfo {pages} {3495} (\bibinfo {year} {2023})},\ \Eprint {https://arxiv.org/abs/2304.06138} {arXiv:2304.06138 [astro-ph.IM]} \BibitemShut {NoStop}%
\bibitem [{\citenamefont {Heinzel}\ and\ \citenamefont {Vitale}(2025)}]{Heinzel:2025ogf}%
  \BibitemOpen
  \bibfield  {author} {\bibinfo {author} {\bibfnamefont {J.}~\bibnamefont {Heinzel}}\ and\ \bibinfo {author} {\bibfnamefont {S.}~\bibnamefont {Vitale}},\ }\Eprint {https://arxiv.org/abs/2509.07221} {arXiv:2509.07221 [astro-ph.HE]}  (\bibinfo {year} {2025})\BibitemShut {NoStop}%
\bibitem [{\citenamefont {Kobayashi}\ \emph {et~al.}(2026)\citenamefont {Kobayashi}, \citenamefont {Iwaya}, \citenamefont {Morisaki}, \citenamefont {Hotokezaka},\ and\ \citenamefont {Kinugawa}}]{Kobayashi:2026lac}%
  \BibitemOpen
  \bibfield  {author} {\bibinfo {author} {\bibfnamefont {K.}~\bibnamefont {Kobayashi}}, \bibinfo {author} {\bibfnamefont {M.}~\bibnamefont {Iwaya}}, \bibinfo {author} {\bibfnamefont {S.}~\bibnamefont {Morisaki}}, \bibinfo {author} {\bibfnamefont {K.}~\bibnamefont {Hotokezaka}},\ and\ \bibinfo {author} {\bibfnamefont {T.}~\bibnamefont {Kinugawa}},\ }\Eprint {https://arxiv.org/abs/2602.12509} {arXiv:2602.12509 [gr-qc]}  (\bibinfo {year} {2026})\BibitemShut {NoStop}%
\bibitem [{\citenamefont {Loredo}(2013)}]{Loredo:2012jm}%
  \BibitemOpen
  \bibfield  {author} {\bibinfo {author} {\bibfnamefont {T.~J.}\ \bibnamefont {Loredo}},\ }in\ \href {https://doi.org/10.1007/978-1-4614-3508-2_2} {\emph {\bibinfo {booktitle} {Astrostatistical Challenges for the New Astronomy}}}\ (\bibinfo  {publisher} {Springer},\ \bibinfo {address} {New York},\ \bibinfo {year} {2013})\ pp.\ \bibinfo {pages} {15--40},\ \Eprint {https://arxiv.org/abs/1208.3036} {arXiv:1208.3036 [astro-ph.IM]} \BibitemShut {NoStop}%
\bibitem [{\citenamefont {Fishbach}\ \emph {et~al.}(2020{\natexlab{b}})\citenamefont {Fishbach}, \citenamefont {Essick},\ and\ \citenamefont {Holz}}]{Fishbach:2020ryj}%
  \BibitemOpen
  \bibfield  {author} {\bibinfo {author} {\bibfnamefont {M.}~\bibnamefont {Fishbach}}, \bibinfo {author} {\bibfnamefont {R.}~\bibnamefont {Essick}},\ and\ \bibinfo {author} {\bibfnamefont {D.~E.}\ \bibnamefont {Holz}},\ }\href {https://doi.org/10.3847/2041-8213/aba7b6} {\bibfield  {journal} {\bibinfo  {journal} {Astrophys. J. Lett.}\ }\textbf {\bibinfo {volume} {899}},\ \bibinfo {pages} {L8} (\bibinfo {year} {2020}{\natexlab{b}})},\ \Eprint {https://arxiv.org/abs/2006.13178} {arXiv:2006.13178 [astro-ph.HE]} \BibitemShut {NoStop}%
\bibitem [{\citenamefont {Essick}\ and\ \citenamefont {Landry}(2020)}]{Essick:2020ghc}%
  \BibitemOpen
  \bibfield  {author} {\bibinfo {author} {\bibfnamefont {R.}~\bibnamefont {Essick}}\ and\ \bibinfo {author} {\bibfnamefont {P.}~\bibnamefont {Landry}},\ }\href {https://doi.org/10.3847/1538-4357/abbd3b} {\bibfield  {journal} {\bibinfo  {journal} {Astrophys. J.}\ }\textbf {\bibinfo {volume} {904}},\ \bibinfo {pages} {80} (\bibinfo {year} {2020})},\ \Eprint {https://arxiv.org/abs/2007.01372} {arXiv:2007.01372 [astro-ph.HE]} \BibitemShut {NoStop}%
\bibitem [{\citenamefont {Romero-Shaw}\ \emph {et~al.}(2022{\natexlab{a}})\citenamefont {Romero-Shaw}, \citenamefont {Thrane},\ and\ \citenamefont {Lasky}}]{Romero-Shaw:2022ctb}%
  \BibitemOpen
  \bibfield  {author} {\bibinfo {author} {\bibfnamefont {I.~M.}\ \bibnamefont {Romero-Shaw}}, \bibinfo {author} {\bibfnamefont {E.}~\bibnamefont {Thrane}},\ and\ \bibinfo {author} {\bibfnamefont {P.~D.}\ \bibnamefont {Lasky}},\ }\href {https://doi.org/10.1017/pasa.2022.24} {\bibfield  {journal} {\bibinfo  {journal} {Publ. Astron. Soc. Aust.}\ }\textbf {\bibinfo {volume} {39}},\ \bibinfo {pages} {e025} (\bibinfo {year} {2022}{\natexlab{a}})},\ \Eprint {https://arxiv.org/abs/2202.05479} {arXiv:2202.05479 [astro-ph.IM]} \BibitemShut {NoStop}%
\bibitem [{\citenamefont {Miller}\ \emph {et~al.}(2026)\citenamefont {Miller}, \citenamefont {Winney}, \citenamefont {Chatziioannou},\ and\ \citenamefont {Meyers}}]{Miller:2026buq}%
  \BibitemOpen
  \bibfield  {author} {\bibinfo {author} {\bibfnamefont {S.~J.}\ \bibnamefont {Miller}}, \bibinfo {author} {\bibfnamefont {S.}~\bibnamefont {Winney}}, \bibinfo {author} {\bibfnamefont {K.}~\bibnamefont {Chatziioannou}},\ and\ \bibinfo {author} {\bibfnamefont {P.~M.}\ \bibnamefont {Meyers}},\ }\Eprint {https://arxiv.org/abs/2604.06090} {arXiv:2604.06090 [gr-qc]}  (\bibinfo {year} {2026})\BibitemShut {NoStop}%
\bibitem [{\citenamefont {Romero-Shaw}\ \emph {et~al.}(2022{\natexlab{b}})\citenamefont {Romero-Shaw}, \citenamefont {Lasky},\ and\ \citenamefont {Thrane}}]{Romero-Shaw:2022xko}%
  \BibitemOpen
  \bibfield  {author} {\bibinfo {author} {\bibfnamefont {I.~M.}\ \bibnamefont {Romero-Shaw}}, \bibinfo {author} {\bibfnamefont {P.~D.}\ \bibnamefont {Lasky}},\ and\ \bibinfo {author} {\bibfnamefont {E.}~\bibnamefont {Thrane}},\ }\href {https://doi.org/10.3847/1538-4357/ac9798} {\bibfield  {journal} {\bibinfo  {journal} {Astrophys. J.}\ }\textbf {\bibinfo {volume} {940}},\ \bibinfo {pages} {171} (\bibinfo {year} {2022}{\natexlab{b}})},\ \Eprint {https://arxiv.org/abs/2206.14695} {arXiv:2206.14695 [astro-ph.HE]} \BibitemShut {NoStop}%
\bibitem [{\citenamefont {Gupte}\ \emph {et~al.}(2025)\citenamefont {Gupte} \emph {et~al.}}]{Gupte:2024jfe}%
  \BibitemOpen
  \bibfield  {author} {\bibinfo {author} {\bibfnamefont {N.}~\bibnamefont {Gupte}} \emph {et~al.},\ }\href {https://doi.org/10.1103/vpyp-nvfp} {\bibfield  {journal} {\bibinfo  {journal} {Phys. Rev. D}\ }\textbf {\bibinfo {volume} {112}},\ \bibinfo {pages} {104045} (\bibinfo {year} {2025})},\ \Eprint {https://arxiv.org/abs/2404.14286} {arXiv:2404.14286 [gr-qc]} \BibitemShut {NoStop}%
\bibitem [{\citenamefont {Morras}\ \emph {et~al.}(2026)\citenamefont {Morras}, \citenamefont {Pratten},\ and\ \citenamefont {Schmidt}}]{Morras:2025xfu}%
  \BibitemOpen
  \bibfield  {author} {\bibinfo {author} {\bibfnamefont {G.}~\bibnamefont {Morras}}, \bibinfo {author} {\bibfnamefont {G.}~\bibnamefont {Pratten}},\ and\ \bibinfo {author} {\bibfnamefont {P.}~\bibnamefont {Schmidt}},\ }\href {https://doi.org/10.3847/2041-8213/ae474c} {\bibfield  {journal} {\bibinfo  {journal} {Astrophys. J. Lett.}\ }\textbf {\bibinfo {volume} {1000}},\ \bibinfo {pages} {L2} (\bibinfo {year} {2026})},\ \Eprint {https://arxiv.org/abs/2503.15393} {arXiv:2503.15393 [astro-ph.HE]} \BibitemShut {NoStop}%
\bibitem [{\citenamefont {Xu}\ \emph {et~al.}(2025)\citenamefont {Xu} \emph {et~al.}}]{Xu:2025ajj}%
  \BibitemOpen
  \bibfield  {author} {\bibinfo {author} {\bibfnamefont {Y.}~\bibnamefont {Xu}} \emph {et~al.},\ }\Eprint {https://arxiv.org/abs/2512.19513} {arXiv:2512.19513 [gr-qc]}  (\bibinfo {year} {2025})\BibitemShut {NoStop}%
\bibitem [{\citenamefont {Pompili}\ \emph {et~al.}(2026)\citenamefont {Pompili}, \citenamefont {Gamboa},\ and\ \citenamefont {Buonanno}}]{Pompili:2026yxq}%
  \BibitemOpen
  \bibfield  {author} {\bibinfo {author} {\bibfnamefont {L.}~\bibnamefont {Pompili}}, \bibinfo {author} {\bibfnamefont {A.}~\bibnamefont {Gamboa}},\ and\ \bibinfo {author} {\bibfnamefont {A.}~\bibnamefont {Buonanno}},\ }\Eprint {https://arxiv.org/abs/2605.28716} {arXiv:2605.28716 [gr-qc]}  (\bibinfo {year} {2026})\BibitemShut {NoStop}%
\bibitem [{\citenamefont {Payne}\ \emph {et~al.}(2023)\citenamefont {Payne}, \citenamefont {Isi}, \citenamefont {Chatziioannou},\ and\ \citenamefont {Farr}}]{Payne:2023kwj}%
  \BibitemOpen
  \bibfield  {author} {\bibinfo {author} {\bibfnamefont {E.}~\bibnamefont {Payne}}, \bibinfo {author} {\bibfnamefont {M.}~\bibnamefont {Isi}}, \bibinfo {author} {\bibfnamefont {K.}~\bibnamefont {Chatziioannou}},\ and\ \bibinfo {author} {\bibfnamefont {W.~M.}\ \bibnamefont {Farr}},\ }\href {https://doi.org/10.1103/PhysRevD.108.124060} {\bibfield  {journal} {\bibinfo  {journal} {Phys. Rev. D}\ }\textbf {\bibinfo {volume} {108}},\ \bibinfo {pages} {124060} (\bibinfo {year} {2023})},\ \Eprint {https://arxiv.org/abs/2309.04528} {arXiv:2309.04528 [gr-qc]} \BibitemShut {NoStop}%
\bibitem [{\citenamefont {Wu}\ and\ \citenamefont {Rong}(2026)}]{Wu:2026qab}%
  \BibitemOpen
  \bibfield  {author} {\bibinfo {author} {\bibfnamefont {T.}~\bibnamefont {Wu}}\ and\ \bibinfo {author} {\bibfnamefont {S.}~\bibnamefont {Rong}},\ }\Eprint {https://arxiv.org/abs/2605.04579} {arXiv:2605.04579 [gr-qc]}  (\bibinfo {year} {2026})\BibitemShut {NoStop}%
\bibitem [{lvk()}]{lvk-data}%
  \BibitemOpen
  \href@noop {} {\bibinfo {title} {\href{https://github.com/mdmould/lvk-data}{https://github.com/mdmould/lvk-data}}}\BibitemShut {NoStop}%
\bibitem [{\citenamefont {{LIGO Scientific Collaboration}}\ and\ \citenamefont {{Virgo Collaboration}}(2021)}]{P2000223}%
  \BibitemOpen
  \bibfield  {author} {\bibinfo {author} {\bibnamefont {{LIGO Scientific Collaboration}}}\ and\ \bibinfo {author} {\bibnamefont {{Virgo Collaboration}}},\ }\href@noop {} {\bibinfo {title} {{LIGO DCC P2000223-v7}, \href{https://dcc.ligo.org/LIGO-P2000223/public}{dcc.ligo.org/LIGO-P2000223/public}}} (\bibinfo {year} {2021})\BibitemShut {NoStop}%
\bibitem [{\citenamefont {{LIGO Scientific Collaboration}}\ and\ \citenamefont {{Virgo Collaboration}}(2022)}]{ligo_scientific_collaboration_and_virgo_2022_6513631}%
  \BibitemOpen
  \bibfield  {author} {\bibinfo {author} {\bibnamefont {{LIGO Scientific Collaboration}}}\ and\ \bibinfo {author} {\bibnamefont {{Virgo Collaboration}}},\ }\href {https://doi.org/10.5281/zenodo.6513631} {10.5281/zenodo.6513631} (\bibinfo {year} {2022})\BibitemShut {NoStop}%
\bibitem [{\citenamefont {{LIGO Scientific Collaboration}}\ \emph {et~al.}(2023)\citenamefont {{LIGO Scientific Collaboration}}, \citenamefont {{Virgo Collaboration}},\ and\ \citenamefont {{KAGRA Collaboration}}}]{ligo_scientific_collaboration_and_virgo_2023_8177023}%
  \BibitemOpen
  \bibfield  {author} {\bibinfo {author} {\bibnamefont {{LIGO Scientific Collaboration}}}, \bibinfo {author} {\bibnamefont {{Virgo Collaboration}}},\ and\ \bibinfo {author} {\bibnamefont {{KAGRA Collaboration}}},\ }\href {https://doi.org/10.5281/zenodo.8177023} {10.5281/zenodo.8177023} (\bibinfo {year} {2023})\BibitemShut {NoStop}%
\bibitem [{\citenamefont {{LIGO Scientific Collaboration}}\ \emph {et~al.}(2025{\natexlab{a}})\citenamefont {{LIGO Scientific Collaboration}}, \citenamefont {{Virgo Collaboration}},\ and\ \citenamefont {{KAGRA Collaboration}}}]{ligo_scientific_collaboration_and_virgo_2025_17014085}%
  \BibitemOpen
  \bibfield  {author} {\bibinfo {author} {\bibnamefont {{LIGO Scientific Collaboration}}}, \bibinfo {author} {\bibnamefont {{Virgo Collaboration}}},\ and\ \bibinfo {author} {\bibnamefont {{KAGRA Collaboration}}},\ }\href {https://doi.org/10.5281/zenodo.17014085} {10.5281/zenodo.17014085} (\bibinfo {year} {2025}{\natexlab{a}})\BibitemShut {NoStop}%
\bibitem [{\citenamefont {{LIGO Scientific Collaboration}}\ \emph {et~al.}(2025{\natexlab{b}})\citenamefont {{LIGO Scientific Collaboration}}, \citenamefont {{Virgo Collaboration}},\ and\ \citenamefont {{KAGRA Collaboration}}}]{ligo_scientific_collaboration_2025_16740128}%
  \BibitemOpen
  \bibfield  {author} {\bibinfo {author} {\bibnamefont {{LIGO Scientific Collaboration}}}, \bibinfo {author} {\bibnamefont {{Virgo Collaboration}}},\ and\ \bibinfo {author} {\bibnamefont {{KAGRA Collaboration}}},\ }\href {https://doi.org/10.5281/zenodo.16740128} {10.5281/zenodo.16740128} (\bibinfo {year} {2025}{\natexlab{b}})\BibitemShut {NoStop}%
\bibitem [{\citenamefont {Taylor}\ and\ \citenamefont {Gerosa}(2018)}]{Taylor:2018iat}%
  \BibitemOpen
  \bibfield  {author} {\bibinfo {author} {\bibfnamefont {S.~R.}\ \bibnamefont {Taylor}}\ and\ \bibinfo {author} {\bibfnamefont {D.}~\bibnamefont {Gerosa}},\ }\href {https://doi.org/10.1103/PhysRevD.98.083017} {\bibfield  {journal} {\bibinfo  {journal} {Phys. Rev. D}\ }\textbf {\bibinfo {volume} {98}},\ \bibinfo {pages} {083017} (\bibinfo {year} {2018})},\ \Eprint {https://arxiv.org/abs/1806.08365} {arXiv:1806.08365 [astro-ph.HE]} \BibitemShut {NoStop}%
\bibitem [{\citenamefont {Essick}\ and\ \citenamefont {Fishbach}(2024)}]{Essick:2023upv}%
  \BibitemOpen
  \bibfield  {author} {\bibinfo {author} {\bibfnamefont {R.}~\bibnamefont {Essick}}\ and\ \bibinfo {author} {\bibfnamefont {M.}~\bibnamefont {Fishbach}},\ }\href {https://doi.org/10.3847/1538-4357/ad1604} {\bibfield  {journal} {\bibinfo  {journal} {Astrophys. J.}\ }\textbf {\bibinfo {volume} {962}},\ \bibinfo {pages} {169} (\bibinfo {year} {2024})},\ \Eprint {https://arxiv.org/abs/2310.02017} {arXiv:2310.02017 [gr-qc]} \BibitemShut {NoStop}%
\bibitem [{\citenamefont {Callister}(2026)}]{Callister:2024cdx}%
  \BibitemOpen
  \bibfield  {author} {\bibinfo {author} {\bibfnamefont {T.~A.}\ \bibnamefont {Callister}},\ }in\ \href {https://doi.org/https://doi.org/10.1016/B978-0-443-21439-4.00092-4} {\emph {\bibinfo {booktitle} {Encyclopedia of Astrophysics (First Edition)}}}\ (\bibinfo  {publisher} {Elsevier},\ \bibinfo {address} {Oxford},\ \bibinfo {year} {2026})\ pp.\ \bibinfo {pages} {546--569},\ \Eprint {https://arxiv.org/abs/2410.19145} {arXiv:2410.19145 [astro-ph.HE]} \BibitemShut {NoStop}%
\bibitem [{\citenamefont {Abbott}\ \emph {et~al.}(2021{\natexlab{c}})\citenamefont {Abbott} \emph {et~al.}}]{LIGOScientific:2019lzm}%
  \BibitemOpen
  \bibfield  {author} {\bibinfo {author} {\bibfnamefont {R.}~\bibnamefont {Abbott}} \emph {et~al.},\ }\href {https://doi.org/10.1016/j.softx.2021.100658} {\bibfield  {journal} {\bibinfo  {journal} {SoftwareX}\ }\textbf {\bibinfo {volume} {13}},\ \bibinfo {pages} {100658} (\bibinfo {year} {2021}{\natexlab{c}})},\ \Eprint {https://arxiv.org/abs/1912.11716} {arXiv:1912.11716 [gr-qc]} \BibitemShut {NoStop}%
\bibitem [{\citenamefont {Abbott}\ \emph {et~al.}(2023{\natexlab{c}})\citenamefont {Abbott} \emph {et~al.}}]{KAGRA:2023pio}%
  \BibitemOpen
  \bibfield  {author} {\bibinfo {author} {\bibfnamefont {R.}~\bibnamefont {Abbott}} \emph {et~al.},\ }\href {https://doi.org/10.3847/1538-4365/acdc9f} {\bibfield  {journal} {\bibinfo  {journal} {Astrophys. J. Supp. S.}\ }\textbf {\bibinfo {volume} {267}},\ \bibinfo {pages} {29} (\bibinfo {year} {2023}{\natexlab{c}})},\ \Eprint {https://arxiv.org/abs/2302.03676} {arXiv:2302.03676 [gr-qc]} \BibitemShut {NoStop}%
\bibitem [{\citenamefont {Abac}\ \emph {et~al.}(2025{\natexlab{f}})\citenamefont {Abac} \emph {et~al.}}]{LIGOScientific:2025snk}%
  \BibitemOpen
  \bibfield  {author} {\bibinfo {author} {\bibfnamefont {A.~G.}\ \bibnamefont {Abac}} \emph {et~al.},\ }\Eprint {https://arxiv.org/abs/2508.18079} {arXiv:2508.18079 [gr-qc]}  (\bibinfo {year} {2025}{\natexlab{f}})\BibitemShut {NoStop}%
\bibitem [{\citenamefont {Varma}\ \emph {et~al.}(2019)\citenamefont {Varma}, \citenamefont {Field}, \citenamefont {Scheel}, \citenamefont {Blackman}, \citenamefont {Gerosa}, \citenamefont {Stein}, \citenamefont {Kidder},\ and\ \citenamefont {Pfeiffer}}]{Varma:2019csw}%
  \BibitemOpen
  \bibfield  {author} {\bibinfo {author} {\bibfnamefont {V.}~\bibnamefont {Varma}}, \bibinfo {author} {\bibfnamefont {S.~E.}\ \bibnamefont {Field}}, \bibinfo {author} {\bibfnamefont {M.~A.}\ \bibnamefont {Scheel}}, \bibinfo {author} {\bibfnamefont {J.}~\bibnamefont {Blackman}}, \bibinfo {author} {\bibfnamefont {D.}~\bibnamefont {Gerosa}}, \bibinfo {author} {\bibfnamefont {L.~C.}\ \bibnamefont {Stein}}, \bibinfo {author} {\bibfnamefont {L.~E.}\ \bibnamefont {Kidder}},\ and\ \bibinfo {author} {\bibfnamefont {H.~P.}\ \bibnamefont {Pfeiffer}},\ }\href {https://doi.org/10.1103/PhysRevResearch.1.033015} {\bibfield  {journal} {\bibinfo  {journal} {Phys. Rev. Research.}\ }\textbf {\bibinfo {volume} {1}},\ \bibinfo {pages} {033015} (\bibinfo {year} {2019})},\ \Eprint {https://arxiv.org/abs/1905.09300} {arXiv:1905.09300 [gr-qc]} \BibitemShut {NoStop}%
\bibitem [{\citenamefont {Pratten}\ \emph {et~al.}(2021)\citenamefont {Pratten} \emph {et~al.}}]{Pratten:2020ceb}%
  \BibitemOpen
  \bibfield  {author} {\bibinfo {author} {\bibfnamefont {G.}~\bibnamefont {Pratten}} \emph {et~al.},\ }\href {https://doi.org/10.1103/PhysRevD.103.104056} {\bibfield  {journal} {\bibinfo  {journal} {Phys. Rev. D}\ }\textbf {\bibinfo {volume} {103}},\ \bibinfo {pages} {104056} (\bibinfo {year} {2021})},\ \Eprint {https://arxiv.org/abs/2004.06503} {arXiv:2004.06503 [gr-qc]} \BibitemShut {NoStop}%
\bibitem [{\citenamefont {Ramos-Buades}\ \emph {et~al.}(2023)\citenamefont {Ramos-Buades}, \citenamefont {Buonanno}, \citenamefont {Estell{\'e}s}, \citenamefont {Khalil}, \citenamefont {Mihaylov}, \citenamefont {Ossokine}, \citenamefont {Pompili},\ and\ \citenamefont {Shiferaw}}]{Ramos-Buades:2023ehm}%
  \BibitemOpen
  \bibfield  {author} {\bibinfo {author} {\bibfnamefont {A.}~\bibnamefont {Ramos-Buades}}, \bibinfo {author} {\bibfnamefont {A.}~\bibnamefont {Buonanno}}, \bibinfo {author} {\bibfnamefont {H.}~\bibnamefont {Estell{\'e}s}}, \bibinfo {author} {\bibfnamefont {M.}~\bibnamefont {Khalil}}, \bibinfo {author} {\bibfnamefont {D.~P.}\ \bibnamefont {Mihaylov}}, \bibinfo {author} {\bibfnamefont {S.}~\bibnamefont {Ossokine}}, \bibinfo {author} {\bibfnamefont {L.}~\bibnamefont {Pompili}},\ and\ \bibinfo {author} {\bibfnamefont {M.}~\bibnamefont {Shiferaw}},\ }\href {https://doi.org/10.1103/PhysRevD.108.124037} {\bibfield  {journal} {\bibinfo  {journal} {Phys. Rev. D}\ }\textbf {\bibinfo {volume} {108}},\ \bibinfo {pages} {124037} (\bibinfo {year} {2023})},\ \Eprint {https://arxiv.org/abs/2303.18046} {arXiv:2303.18046 [gr-qc]} \BibitemShut {NoStop}%
\bibitem [{\citenamefont {Ossokine}\ \emph {et~al.}(2020)\citenamefont {Ossokine} \emph {et~al.}}]{Ossokine:2020kjp}%
  \BibitemOpen
  \bibfield  {author} {\bibinfo {author} {\bibfnamefont {S.}~\bibnamefont {Ossokine}} \emph {et~al.},\ }\href {https://doi.org/10.1103/PhysRevD.102.044055} {\bibfield  {journal} {\bibinfo  {journal} {Phys. Rev. D}\ }\textbf {\bibinfo {volume} {102}},\ \bibinfo {pages} {044055} (\bibinfo {year} {2020})},\ \Eprint {https://arxiv.org/abs/2004.09442} {arXiv:2004.09442 [gr-qc]} \BibitemShut {NoStop}%
\bibitem [{gwa()}]{gwax}%
  \BibitemOpen
  \href@noop {} {\bibinfo {title} {\href{https://github.com/mdmould/gwax}{https://github.com/mdmould/gwax}}}\BibitemShut {NoStop}%
\bibitem [{\citenamefont {Essick}\ \emph {et~al.}(2025)\citenamefont {Essick} \emph {et~al.}}]{Essick:2025zed}%
  \BibitemOpen
  \bibfield  {author} {\bibinfo {author} {\bibfnamefont {R.}~\bibnamefont {Essick}} \emph {et~al.},\ }\href {https://doi.org/10.1103/44x3-hv3y} {\bibfield  {journal} {\bibinfo  {journal} {Phys. Rev. D}\ }\textbf {\bibinfo {volume} {112}},\ \bibinfo {pages} {102001} (\bibinfo {year} {2025})},\ \Eprint {https://arxiv.org/abs/2508.10638} {arXiv:2508.10638 [gr-qc]} \BibitemShut {NoStop}%
\bibitem [{\citenamefont {Tiwari}(2018)}]{Tiwari:2017ndi}%
  \BibitemOpen
  \bibfield  {author} {\bibinfo {author} {\bibfnamefont {V.}~\bibnamefont {Tiwari}},\ }\href {https://doi.org/10.1088/1361-6382/aac89d} {\bibfield  {journal} {\bibinfo  {journal} {Class. Quantum Grav.}\ }\textbf {\bibinfo {volume} {35}},\ \bibinfo {pages} {145009} (\bibinfo {year} {2018})},\ \Eprint {https://arxiv.org/abs/1712.00482} {arXiv:1712.00482 [astro-ph.HE]} \BibitemShut {NoStop}%
\bibitem [{\citenamefont {Ashton}\ \emph {et~al.}(2019)\citenamefont {Ashton} \emph {et~al.}}]{Ashton:2018jfp}%
  \BibitemOpen
  \bibfield  {author} {\bibinfo {author} {\bibfnamefont {G.}~\bibnamefont {Ashton}} \emph {et~al.},\ }\href {https://doi.org/10.3847/1538-4365/ab06fc} {\bibfield  {journal} {\bibinfo  {journal} {Astrophys. J. Supp. S.}\ }\textbf {\bibinfo {volume} {241}},\ \bibinfo {pages} {27} (\bibinfo {year} {2019})},\ \Eprint {https://arxiv.org/abs/1811.02042} {arXiv:1811.02042 [astro-ph.IM]} \BibitemShut {NoStop}%
\bibitem [{\citenamefont {Speagle}(2020)}]{Speagle:2019ivv}%
  \BibitemOpen
  \bibfield  {author} {\bibinfo {author} {\bibfnamefont {J.~S.}\ \bibnamefont {Speagle}},\ }\href {https://doi.org/10.1093/mnras/staa278} {\bibfield  {journal} {\bibinfo  {journal} {Mon. Not. R. Astron. Soc.}\ }\textbf {\bibinfo {volume} {493}},\ \bibinfo {pages} {3132} (\bibinfo {year} {2020})},\ \Eprint {https://arxiv.org/abs/1904.02180} {arXiv:1904.02180 [astro-ph.IM]} \BibitemShut {NoStop}%
\bibitem [{\citenamefont {Fishbach}\ \emph {et~al.}(2018)\citenamefont {Fishbach}, \citenamefont {Holz},\ and\ \citenamefont {Farr}}]{Fishbach:2018edt}%
  \BibitemOpen
  \bibfield  {author} {\bibinfo {author} {\bibfnamefont {M.}~\bibnamefont {Fishbach}}, \bibinfo {author} {\bibfnamefont {D.~E.}\ \bibnamefont {Holz}},\ and\ \bibinfo {author} {\bibfnamefont {W.~M.}\ \bibnamefont {Farr}},\ }\href {https://doi.org/10.3847/2041-8213/aad800} {\bibfield  {journal} {\bibinfo  {journal} {Astrophys. J. Lett.}\ }\textbf {\bibinfo {volume} {863}},\ \bibinfo {pages} {L41} (\bibinfo {year} {2018})},\ \Eprint {https://arxiv.org/abs/1805.10270} {arXiv:1805.10270 [astro-ph.HE]} \BibitemShut {NoStop}%
\bibitem [{\citenamefont {Ade}\ \emph {et~al.}(2016)\citenamefont {Ade} \emph {et~al.}}]{Planck:2015fie}%
  \BibitemOpen
  \bibfield  {author} {\bibinfo {author} {\bibfnamefont {P.~A.~R.}\ \bibnamefont {Ade}} \emph {et~al.},\ }\href {https://doi.org/10.1051/0004-6361/201525830} {\bibfield  {journal} {\bibinfo  {journal} {Astron. Astrophys.}\ }\textbf {\bibinfo {volume} {594}},\ \bibinfo {pages} {A13} (\bibinfo {year} {2016})},\ \Eprint {https://arxiv.org/abs/1502.01589} {arXiv:1502.01589 [astro-ph.CO]} \BibitemShut {NoStop}%
\bibitem [{\citenamefont {De~Renzis}\ \emph {et~al.}(2025)\citenamefont {De~Renzis}, \citenamefont {Iacovelli}, \citenamefont {Gerosa}, \citenamefont {Mancarella},\ and\ \citenamefont {Pacilio}}]{DeRenzis:2024dvx}%
  \BibitemOpen
  \bibfield  {author} {\bibinfo {author} {\bibfnamefont {V.}~\bibnamefont {De~Renzis}}, \bibinfo {author} {\bibfnamefont {F.}~\bibnamefont {Iacovelli}}, \bibinfo {author} {\bibfnamefont {D.}~\bibnamefont {Gerosa}}, \bibinfo {author} {\bibfnamefont {M.}~\bibnamefont {Mancarella}},\ and\ \bibinfo {author} {\bibfnamefont {C.}~\bibnamefont {Pacilio}},\ }\href {https://doi.org/10.1103/PhysRevD.111.044048} {\bibfield  {journal} {\bibinfo  {journal} {Phys. Rev. D}\ }\textbf {\bibinfo {volume} {111}},\ \bibinfo {pages} {044048} (\bibinfo {year} {2025})},\ \Eprint {https://arxiv.org/abs/2410.17325} {arXiv:2410.17325 [astro-ph.HE]} \BibitemShut {NoStop}%
\end{thebibliography}%

\end{document}